%% file: draft_v8.0.tex
\RequirePackage[mathlines]{lineno} 
\documentclass[aps,prd,twocolumn,showpacs,amsmath,amssymb]{revtex4-1}
\usepackage{epsfig}
\usepackage{graphicx}
\usepackage{dcolumn}
\usepackage{bm}
\usepackage{overpic}
\usepackage{subfigure}
\usepackage{float}
\usepackage{color}
\usepackage{amsmath}
\usepackage{mathcomp}
\usepackage{multirow}
\usepackage{rotating}

\begin{document}
\normalsize
\parskip=5pt plus 1pt minus 1pt

\title{ \boldmath Confirmation of a charged charmoniumlike state $Z_c(3885)^{\mp}$ in
$e^+e^-\to\pi^{\pm}(D\bar{D}^*)^\mp$ with double $D$ tag}

\input{authors_Sep2015}

\vspace{4cm}

\date{\today}

\begin{abstract}
We present a study of the process $e^+e^-\to\pi^{\pm}(D\bar{D}^*)^{\mp}$
using data samples of 1092~pb$^{-1}$ at $\sqrt{s}=4.23$~GeV
and 826~pb$^{-1}$ at $\sqrt{s}=4.26$~GeV collected with the BESIII detector at the BEPCII storage ring.
With full reconstruction of the $D$ meson pair and the bachelor $\pi^{\pm}$ in the final state,
we confirm the existence of the charged structure $Z_c(3885)^{\mp}$ in the $(D\bar{D}^*)^{\mp}$ system in the two
isospin processes $e^+e^-\to\pi^+D^0D^{*-}$ and $e^+e^-\to\pi^+D^-D^{*0}$.
By performing a simultaneous fit, the statistical significance of $Zc(3885)^{\mp}$ signal
is determined to be greater than 10$\sigma$, and its pole mass and width are measured to be
$M_{\rm{pole}}$=(3881.7$\pm$1.6(stat.)$\pm$1.6(syst.))~MeV/$c^2$
and $\Gamma_{\rm{pole}}$=(26.6$\pm$2.0(stat.)$\pm$2.1(syst.))~MeV, respectively.
The Born cross section times the $(D\bar{D}^*)^{\mp}$ branching fraction
($\sigma(e^+e^-\to\pi^{\pm}Z_{c}(3885)^{\mp}) \times Br(Z_{c}(3885)^{\mp}\to(D\bar{D}^*)^{\mp})$)
is measured to be $(141.6\pm7.9(\text{stat.})\pm12.3(\text{syst.}))~\text{pb}$ at $\sqrt{s}=4.23$~GeV
and $(108.4\pm6.9(\text{stat.})\pm8.8(\text{syst.}))~\text{pb}$ at $\sqrt{s}=4.26$~GeV.
The polar angular distribution of the $\pi^{\pm}$-$Z_c(3885)^{\mp}$ system
is consistent with the expectation of a quantum number assignment of $J^P=1^+$ for $Z_c(3885)^{\mp}$.
\end{abstract}


\maketitle

\input {introduction}
\input {BESIII}
\input {selection}
\input {signal}
\input {syserr}
\input {summary}
\input {acknowledgement}

\input {biblio}
\input {appA}
\input {appB}
\end{document}

%% file: authors_Sep2015.tex
\author{
  \begin{small}
    \begin{center}
      M.~Ablikim$^{1}$, M.~N.~Achasov$^{9,f}$, X.~C.~Ai$^{1}$,
      O.~Albayrak$^{5}$, M.~Albrecht$^{4}$, D.~J.~Ambrose$^{44}$,
      A.~Amoroso$^{49A,49C}$, F.~F.~An$^{1}$, Q.~An$^{46,a}$,
      J.~Z.~Bai$^{1}$
% ...
      \\
      \vspace{0.2cm}
      (BESIII Collaboration)\\
      \vspace{0.2cm} {\it
        $^{1}$ Institute of High Energy Physics, Beijing 100049, People's Republic of China\\
% ...
$^{4}$ Bochum Ruhr-University, D-44780 Bochum, Germany\\
        $^{5}$ Carnegie Mellon University, Pittsburgh, Pennsylvania 15213, USA\\
% ...
$^{9}$ G.I. Budker Institute of Nuclear Physics SB RAS (BINP), Novosibirsk 630090, Russia\\
% ...
$^{44}$ University of Rochester, Rochester, New York 14627, USA\\
% ...
$^{46}$ University of Science and Technology of China, Hefei 230026, People's Republic of China\\
% ...
$^{49}$ (A)University of Turin, I-10125, Turin, Italy; (B)University of Eastern Piedmont, I-15121, Alessandria, Italy; (C)INFN, I-10125, Turin, Italy\\
% ...
$^{a}$ Also at State Key Laboratory of Particle Detection and Electronics, Beijing 100049, Hefei 230026, People's Republic of China\\
% ...
$^{f}$ Also at the Novosibirsk State University, Novosibirsk, 630090, Russia\\
% ...
      }
    \end{center}
    \vspace{2cm}
  \end{small}
}

\affiliation{}


%% file: introduction.tex
\section{Introduction}
\label{sec:introduction}
\vspace{-0.4cm}
The $Y(4260)$ was first observed by BaBar in the initial-state-radiation (ISR) process
$e^+e^- \to \gamma_{ISR}\pi^+\pi^-J/\psi$~\cite{PRL95-142001}.
This observation was subsequently confirmed by CLEO~\cite{PRD74-091104} and Belle~\cite{PRL99-182004}.
Unlike other charmonium states, such as $\psi(4040)$, $\psi(4160)$ and $\psi(4415)$,
$Y(4260)$ does not have a natural place within the quark model of charmonium~\cite{PRD72-054026}.
Many theoretical interpretations have been proposed to understand the underlying structure of
$Y(4260)$~\cite{PLB625-212-2005, PLB631-164-2005, PLB628-215-2005}, more precise experiments are
necessary to give a decisive conclusion.

In recent years, a common pattern has been observed for the charmoniumlike
states in the systems $\pi J/\psi$, $\pi \psi^\prime$, $\pi h_c$ and $\pi \chi_c$
as well as in pairs of charmed mesons $D\bar{D}^{*}$ and $D^{*}\bar{D}^{*}$.
Belle observed some charged structures called $Z(4430)^{\pm}$
in the $\pi^{\pm}\psi^\prime$ system~\cite{PRL100-142001, PRD80-031104, PRD88-074026},
and $Z_1(4050)^{\pm}$ and $Z_2(4250)^{\pm}$ in the $\pi^{\pm}\chi_{c_1}$
invariant mass spectra~\cite{PRD78-072004} in $B$ meson decays.
The $Z(4430)^{\pm}$ has recently been confirmed by LHCb~\cite{PRL112-222002} in the $\pi^{\pm}\psi^\prime$ system.
However, neither $Z_1(4050)^{\pm}$ nor $Z_2(4250)^{\pm}$ are found to be significant
in BaBar data~\cite{PRD79-112001, PRD85-052003}.
BESIII~\cite{PRL110-252001} and Belle~\cite{PRL110-252002} observed the $Z_c(3900)^{\pm}$ in the $\pi^{\pm}J/\psi$
invariant mass distribution in a study of $e^+e^-\to \pi^+\pi^-J/\psi$;
this observation was confirmed with CLEOc data at $\sqrt{s}$=4.17~GeV~\cite{PLB727-366-2013}.
More recently, BESIII has reported the observations of
the $Z_c(3900)^0$ in the $\pi^0 J/\psi$ system~\cite{arxiv:150606018},
$Z_c(4020)$ in the $\pi h_c$ system~\cite{PRL111-242001, PRL113-212002},
$Z_c(4025)$ in the $D^{*}\bar{D}^{*}$ system~\cite{PRL112-132001, arxiv:150702404},
and $Z_c(3885)^{\pm}$ in the $(D\bar{D}^{*})^{\pm}$ system~\cite{PRL112-022001}.
It is interesting to note that all these states lie close to
the threshold of some charm meson pair systems and some of them even have overlapping widths.
It is therefore important to obtain more experimental information to improve the understanding of all these states.

In a previous paper by BESIII~\cite{PRL112-022001},
a structure called $Z_c(3885)^{\pm}$ was observed in the study of $e^+e^-\to \pi^+ D^0 D^{*-}$ ($D^0 \to K^-\pi^+$)
and $e^+e^-\to \pi^+ D^- D^{*0}$ ($D^- \to K^+\pi^-\pi^-$)
using a 525~pb$^{-1}$ subset of the data sample collected around $\sqrt{s}=4.26$~GeV.
That study employs a partial reconstruction technique by reconstructing one final-state $D$ meson
and the bachelor $\pi$ coming directly from $e^+e^-$ decay
(\textquotedblleft single $D$ tag\textquotedblright or ST) and inferring the presence of the $\bar{D}^*$
from energy-momentum conservation.
In this analysis, we present a combined study of the processes $e^+e^-\to \pi^+D^0D^{*-}$ ($\pi^+D^0\bar{D}^0$-tagged)
and $e^+e^-\to \pi^+ D^-D^{*0}$ ($\pi^+D^-D^0$-tagged) 
using data samples of 1092~pb$^{-1}$ at $\sqrt{s}$=4.23~GeV and 826~pb$^{-1}$
at $\sqrt{s}$=4.26~GeV~\cite{arxiv:150303408}
collected with the BESIII detector at the BEPCII storage ring
(charge conjugated processes are included throughout this paper).
We reconstruct the bachelor $\pi^+$ and the $D$ meson pair
(\textquotedblleft double $D$ tag\textquotedblright or DT) in the final state.
Because the $\pi$ from $D^{*-}$ and $D^{*0}$ decays has low momentum, it is difficult to reconstruct directly. 
We denote it as the \textquotedblleft missing $\pi$\textquotedblright{} and infer its presence using energy-momentum conservation.
The $D^0$ mesons are reconstructed in four decay modes and the $D^-$ mesons in six decay modes.
The double $D$ tag technique allows the use of more $D$ decay modes and effectively suppresses backgrounds.

%% file: BESIII.tex
\section{Experiment And Data Sample} 
\label{sec:BESIII}
\vspace{-0.4cm}
The BESIII detector is described in detail elsewhere~\cite{bes3-dector}.
It has an effective geometrical acceptance of 93\% of 4$\pi$.
It consists of a small-cell, helium-based (40\% He, 60\% C$_3$H$_{8}$) main drift chamber (MDC),
a plastic scintillator time-of-flight system (TOF),
a CsI(TI) electromagnetic calorimeter (EMC) and
a muon system (MUC) containing resistive plate chambers (RPC) in the iron return yoke of a 1~T superconducting solenoid.
The momentum resolution for charged tracks is 0.5\% at a momentum of 1~GeV/$c$.
Charged particle identification (PID) is accomplished by
combining the energy loss ($dE/dx$) measurements in the MDC and flight times in the TOF.
The photon energy resolution at 1~GeV is 2.5\% in the barrel and 5\% in the end caps.

The GEANT4-based~\cite{GEANT4-Col,GEANT4-Allison} Monte Carlo (MC) simulation software BOOST~\cite{dengzy}
includes the geometric and material description of the BESIII detectors,
the detector response and digitization models,
as well as the tracking of the detector running conditions and performance.
It is used to optimize the selection criteria, to evaluate the signal efficiency
and mass resolution, and to estimate the physics backgrounds.
The physics backgrounds are studied using a generic MC sample which consists of the production of the $Y(4260)$ state
and its exclusive decays, the process $e^+e^- \to (\pi)D^{(*)}\bar{D}^{(*)}$, the production of ISR photons to low mass $\psi$
states, and QED processes.
The $Y(4260)$ resonance, ISR production of the vector charmonium states, and QED events are generated by KKMC~\cite{kkmc}.
The known decay modes are generated by EVTGEN~\cite{evtgen,simulation-RGPing} with branching ratios being set to
world average values from the Particle Data Group (PDG)~\cite{pdg}, and
the remaining unknown decay modes are generated by LUNDCHARM~\cite{PRD62-034003}.
In addition, exclusive MC samples for the process $e^+e^- \to D_{J}\bar{D}^*, D_{J}\to D^{(*)}\pi(\pi)$
are generated to study the possible background contributions from neutral and charged highly excited $D$ states
(denoted as $D_J$, where $J$ is the spin of the meson),
such as $D^{*}_0(2400)$, $D_1(2420)$, $D_1(2430)$ and $D_2^*(2460)$.
To estimate the signal efficiency and to optimize the selection criteria, 
we generate a signal MC sample for the process $e^+e^- \to \pi^+Z_c(3885)^-~(Z_c(3885)^- \to (D\bar{D}^*)^-)$
and a phase space MC sample (PHSP MC) for the process $e^+e^- \to \pi^+(D\bar{D}^*)^-$.
Here the spin and parity of the $Z_c(3885)^{-}$ state are assumed to be $1^{+}$, which is consistent with our observation.

%% file: selection.tex
\section{Event Selection And Background Analysis}
\label{sec:selection}
\vspace{-0.4cm}
Charged tracks are reconstructed in the MDC. For each good charged track, the polar angle must satisfy
$|\cos\theta|<0.93$, and its point of closest approach to the interaction point must be within 10~cm
in the beam direction and within 1~cm in the plane perpendicular to the beam direction.
To assign a particle hypothesis to the charged track, $dE/dx$ and TOF information
are combined to form a probability $\mathrm{Prob}(K)$ ($\mathrm{Prob}(\pi)$).
A track is identified as a $K$ ($\pi$) when $\mathrm{Prob}(K) > \mathrm{Prob}(\pi)$ ($\mathrm{Prob}(\pi) > \mathrm{Prob}(K)$).
Tracks used in reconstructing $K^0_S$ decays are exempted from these requirements.

Photon candidates are reconstructed by clustering EMC crystal energies.
For each photon candidate, the energy deposit in the EMC barrel region ($|\cos\theta|<0.8$) is required to be greater than 25~MeV
and in the EMC endcap region ($0.84 < |\cos\theta| < 0.92$) greater than 50~MeV.
To eliminate showers from charged particles,
the angle between the photon and the nearest charged track is
required to be greater than $20\tcdegree$.
Timing requirements are used to suppress electronic noise
and energy deposits in the EMC unrelated to the event.

We reconstruct $\pi^0$ candidates from pairs of photons with an invariant mass
in the range $0.115<M_{\gamma\gamma}<0.150$~MeV/$c^2$. A one-constraint (1C) kinematic fit
is performed to improve the energy resolution, with $M_{\gamma\gamma}$ being constrained to the known $\pi^0$ mass from PDG~\cite{pdg}.

$K^0_S$ candidates are reconstructed from pairs of oppositely charged tracks which satisfy $|\cos\theta|<0.93$
for the polar angle and the distance of the track to the interaction point in the beam direction within 20~cm.
For each candidate, we perform a vertex fit constraining the charged tracks to a common decay vertex and
use the corrected track parameters to calculate the invariant mass which must be in the range $0.487<M_{\pi^+\pi^-}<0.511$~GeV/$c^2$.
To reject random $\pi^+\pi^-$ combinations, a secondary-vertex fitting algorithm is employed to impose a kinematic
constraint between the production and decay vertices~\cite{CPC-33-428}.

\begin{figure*}[hbtp]
\centering
\epsfig{width=0.48\textwidth,clip=true,file=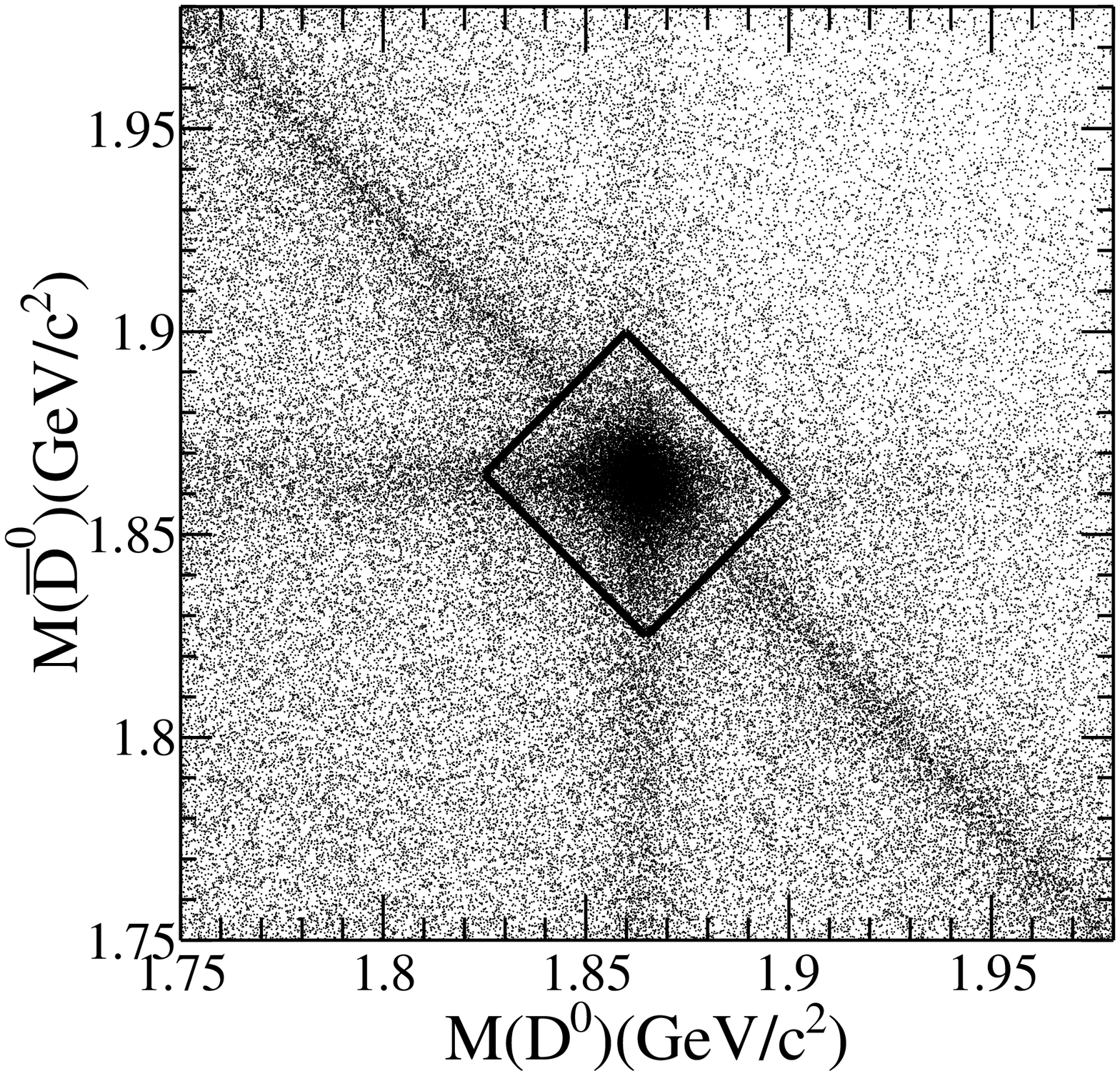}
\epsfig{width=0.48\textwidth,clip=true,file=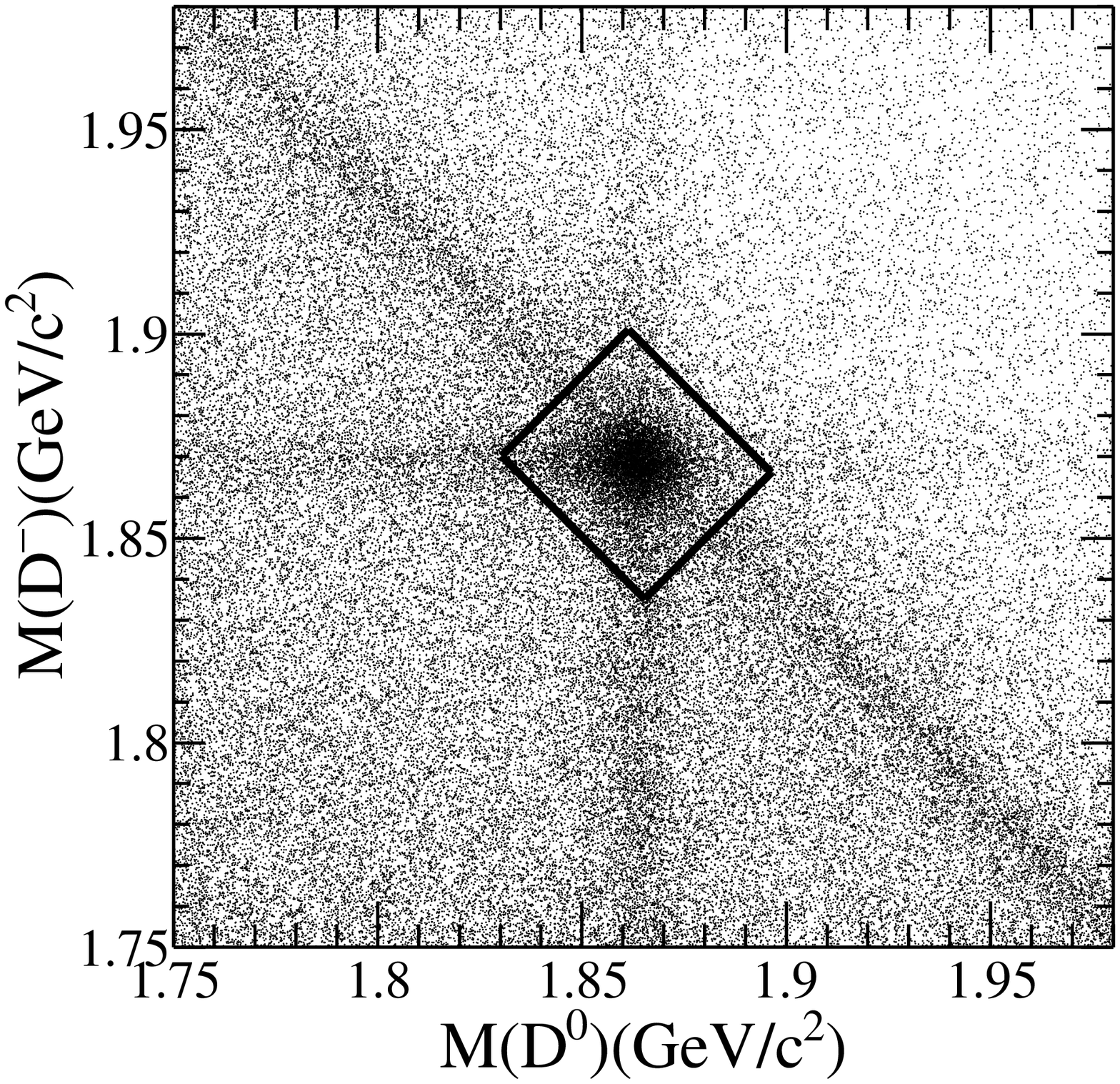}
\caption{Masses of the $\bar{D}$ and $D$ candidates for all DT modes at $\sqrt{s}$=4.26~GeV.
The vertical (horizontal) bands centered at $M(D)$ ($M(\bar{D})$)
contain the DT candidates in which the $D$ ($\bar{D}$) candidate was reconstructed correctly, but the $\bar{D}$ ($D$) was not.
The diagonal bands contain the \textquotedblleft mis-reconstructed\textquotedblright $D\bar{D}$ candidates
(all of the $\bar{D}$ and $D$ final states were reconstructed,
but one or more final states from the $D$ were interchanged with corresponding particles from the $\bar{D}$).
Other combinatorial candidates with minimum $\Delta\hat{M}$ also spread along the diagonal.
The left plot shows $M(\bar{D}^0)$ versus $M(D^0)$, while the right plot shows $M(D^-)$ versus $M(D^0)$.
The solid rectangles show the signal regions.}
\label{DTag_mass}
\end{figure*}

The selected $\pi^{\pm}$, $K^{\pm}$, $\pi^0$ and $K^0_S$ are used to
reconstruct $D$ meson candidates for the $D^0\bar{D}^0$ and $D^- D^0$ double tag.
The $D^0$ candidates are reconstructed in four final states: $K^-\pi^+$, $K^-\pi^+\pi^0$,
$K^-\pi^+\pi^+\pi^-$ and $K^-\pi^+\pi^+\pi^-\pi^0$ (in the following labeled as 0, 1, 2, and 3, respectively), and the $D^-$ candidates in
six final states: $K^+\pi^-\pi^-$, $K^+\pi^-\pi^-\pi^0$, $K^0_S\pi^-$, $K^0_S\pi^-\pi^0$,
$K^0_S\pi^+\pi^-\pi^-$ and $K^+K^-\pi^-$ (labeled as $A,~B,~C,~D,~E$ and $F$, respectively).
If there is more than one candidate per possible DT mode, the candidate with the minimum $\Delta\hat{M}$ is chosen,
where $\Delta\hat{M}$ is the difference between the average mass $\hat{M} =[M(D)+M(\bar{D})]/2$
and $[M_\mathrm{PDG}(D)+M_\mathrm{PDG}(\bar{D})]/2$ ($M_\mathrm{PDG}(D)$ and $M_\mathrm{PDG}(\bar{D})$ are the $D$ mass and $\bar{D}$
mass from PDG~\cite{pdg}, respectively).
Figure~\ref{DTag_mass} shows the distributions of $M(\bar{D})$ versus $M(D)$ for all DT candidates at $\sqrt{s}$=4.26~GeV.
The combinatorial background tends to have structure in $\Delta\hat{M}$
but is flat in the mass difference $\Delta M = M(D)-M(\bar{D})$.
The signal region in the $M(\bar{D})$ versus $M(D)$ plane is defined as
$-20<\Delta\hat{M}<15$~MeV/$c^2$ ($-17<\Delta\hat{M}<14$~MeV/$c^2$)
and $|\Delta M| < 40$~MeV/$c^2$ ($|\Delta M| < 35$~MeV/$c^2$)
for $D^0\bar{D}^0$ ($D^-D^0$) candidates.

\begin{figure*}[!hbtp]
\centering
\begin{minipage}[b]{0.48\textwidth}
\epsfig{width=0.98\textwidth,clip=true,file=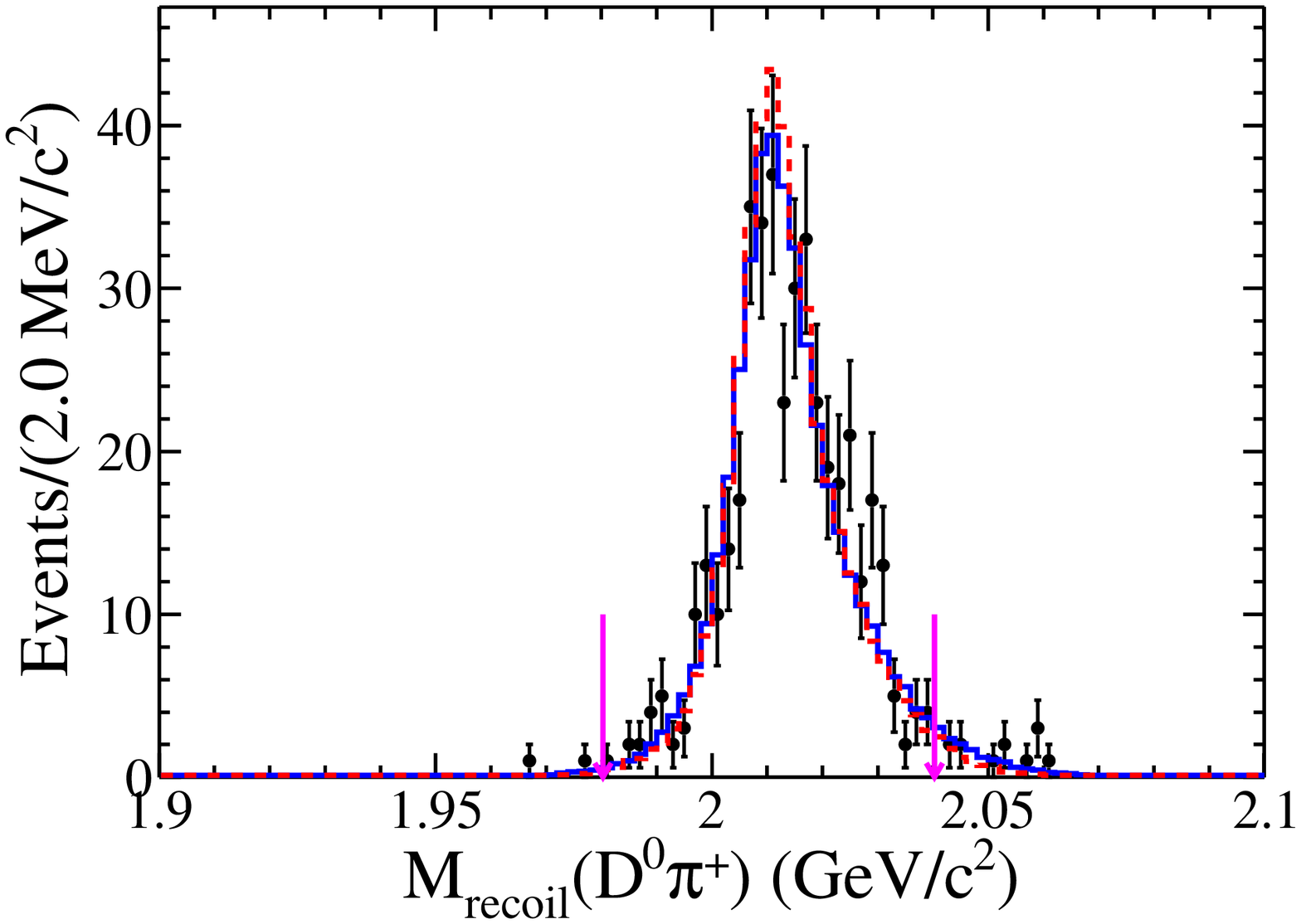}
\put(-50,120){\bf\boldmath (a)}
\end{minipage}
\begin{minipage}[b]{0.48\textwidth}
\epsfig{width=0.98\textwidth,clip=true,file=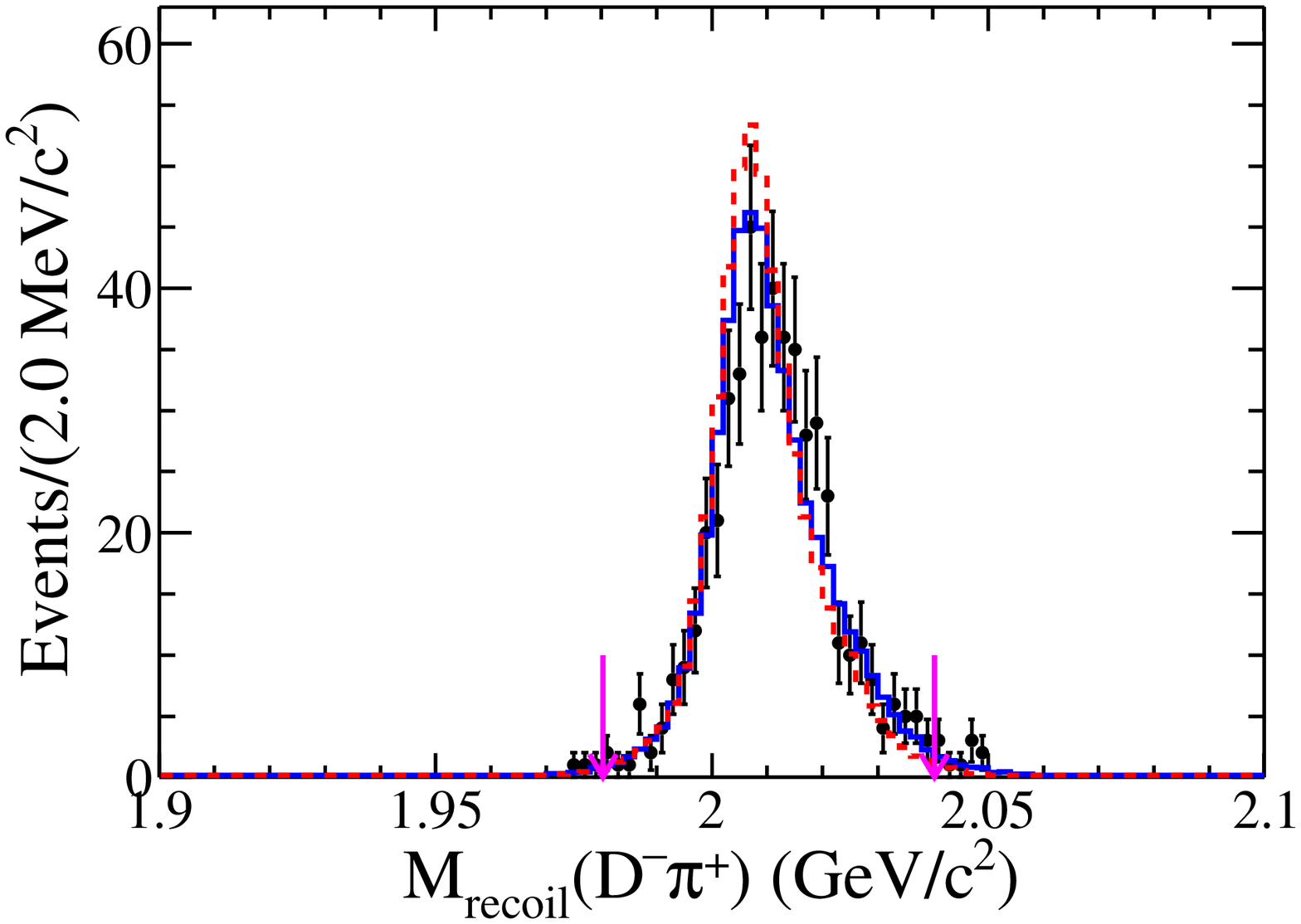}
\put(-50,120){\bf\boldmath (b)}
\end{minipage}
\caption{The $M_\mathrm{recoil}(D\pi)$ distributions for (a) $\pi^+D^0\bar{D}^0$-tagged
events and (b) $\pi^+D^-D^{0}$-tagged events at $\sqrt{s}$=4.26~GeV.
The dots with error bars are data.
The dashed (red) and solid (blue) lines are signal MC and PHSP MC, respectively.
The arrows (pink) indicate nominal selection criteria.}
\label{recoil_Dpi}
\end{figure*}

To reconstruct the bachelor $\pi^+$,
at least one additional good charged track which is not among the decay products of the $D$ candidates is required.
To reduce background and improve the mass resolution,
we perform a four-constraint (4C) kinematic fit to the selected events.
It imposes momentum and energy conservation,
constrains the invariant mass of $D$ ($\bar{D}$) candidates to $M_\mathrm{PDG}(D)$ ($M_\mathrm{PDG}(\bar{D})$),
and constrains the invariant mass formed from the missing $\pi$ and the corresponding $D$ candidate
to $M_\mathrm{PDG}(D^*)$~\cite{pdg}.
This gives a total of 7 constraints.
The missing $\pi$ three-momentum needs to be determined, so we are left with a four-constraint fit.
The $\chi^2$ of the 4C kinematic fit ($\chi^2_{4C}$) is required to be less than 100.
If there are multiple candidates in an event, we choose the one with minimum $\chi^2_{4C}$.
To suppress the background process $e^+e^- \to D^{*}\bar{D}^{*}$,
we require $M(\pi^+D^0) > 2.03$~GeV/$c^2$ ($M(\pi^+D^-) > 2.08$~GeV/$c^2$) for
$\pi^+D^0\bar{D}^0$-tagged ($\pi^+D^-D^0$-tagged) events.
We define the reconstructed $D\pi$ recoil mass $M_\mathrm{recoil}(D\pi)$ via
$M_\mathrm{recoil}(D\pi)^2c^4 =(E_{\rm{cm}}-E_D-E_{\pi})^2 -
|\pmb{\rm{p}}_{cm}-\pmb{\rm{p}}_D-\pmb{\rm{p}}_{\pi}|^2c^2$,
where ($E_{\rm{cm}}$, $\pmb{\rm{p}}_{cm}$), ($E_D$, $\pmb{\rm{p}}_D$) and ($E_{\pi}$, $\pmb{\rm{p}}_{\pi}$)
are the four momentum of the $e^+e^-$ system, $D$ and $\pi$ in the $e^+e^-$ rest frame, respectively.
Figure~\ref{recoil_Dpi} shows the $M_\mathrm{recoil}(D\pi)$ distributions at $\sqrt{s}$=4.26~GeV
after all of the above selection criteria.
The results of signal MC and PHSP MC are provided
to verify the signal processes and optimize the selection criteria.
A study of generic MC sample shows that very few background events which can satisfy the above requirements.

\begin{figure*}[htbp]
\centering
\begin{minipage}[b]{0.45\textwidth}
\epsfig{width=0.98\textwidth,clip=true,file=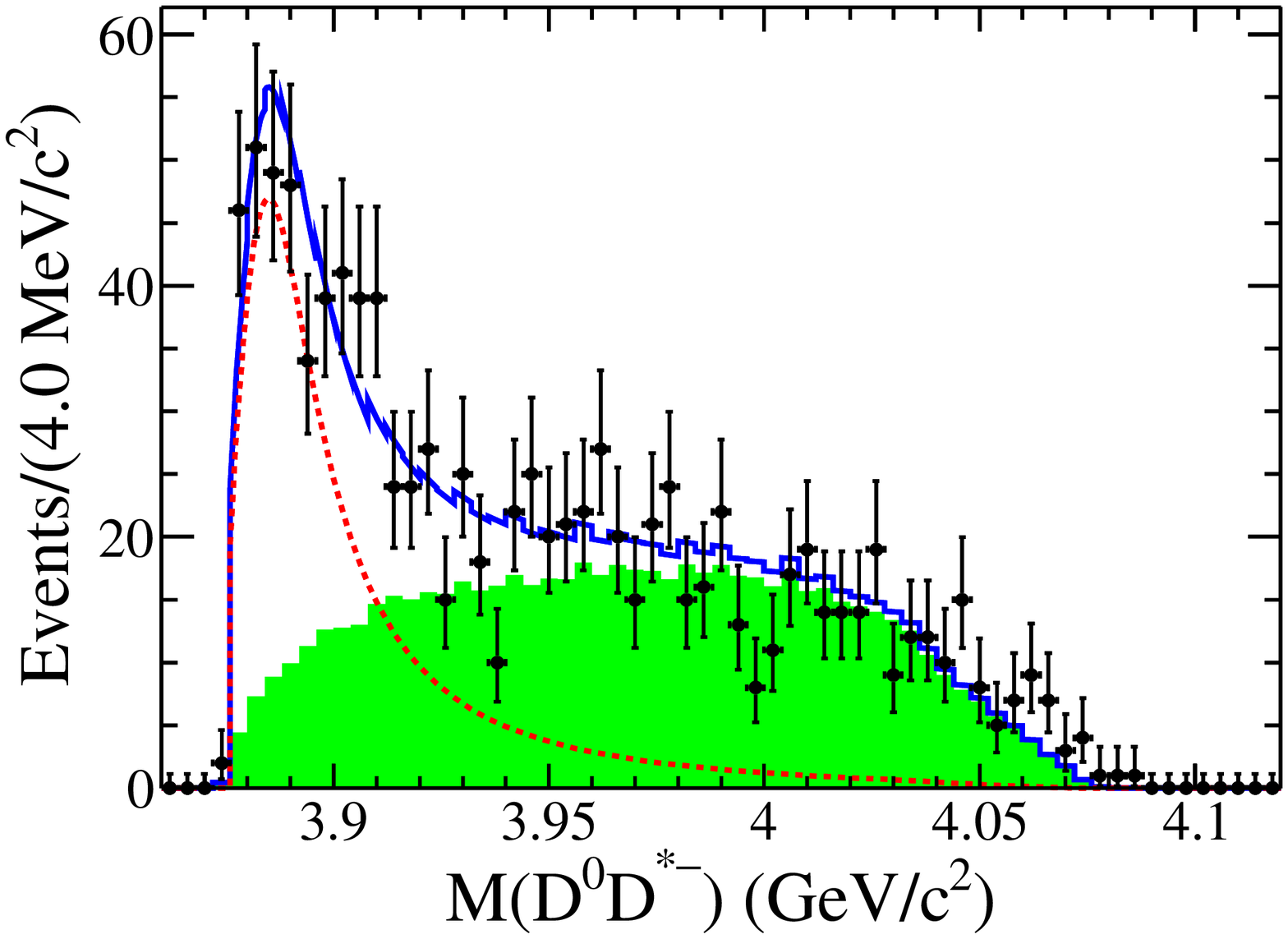}
\put(-50,120){\bf\boldmath (a)}
\end{minipage}
\begin{minipage}[b]{0.45\textwidth}
\epsfig{width=0.98\textwidth,clip=true,file=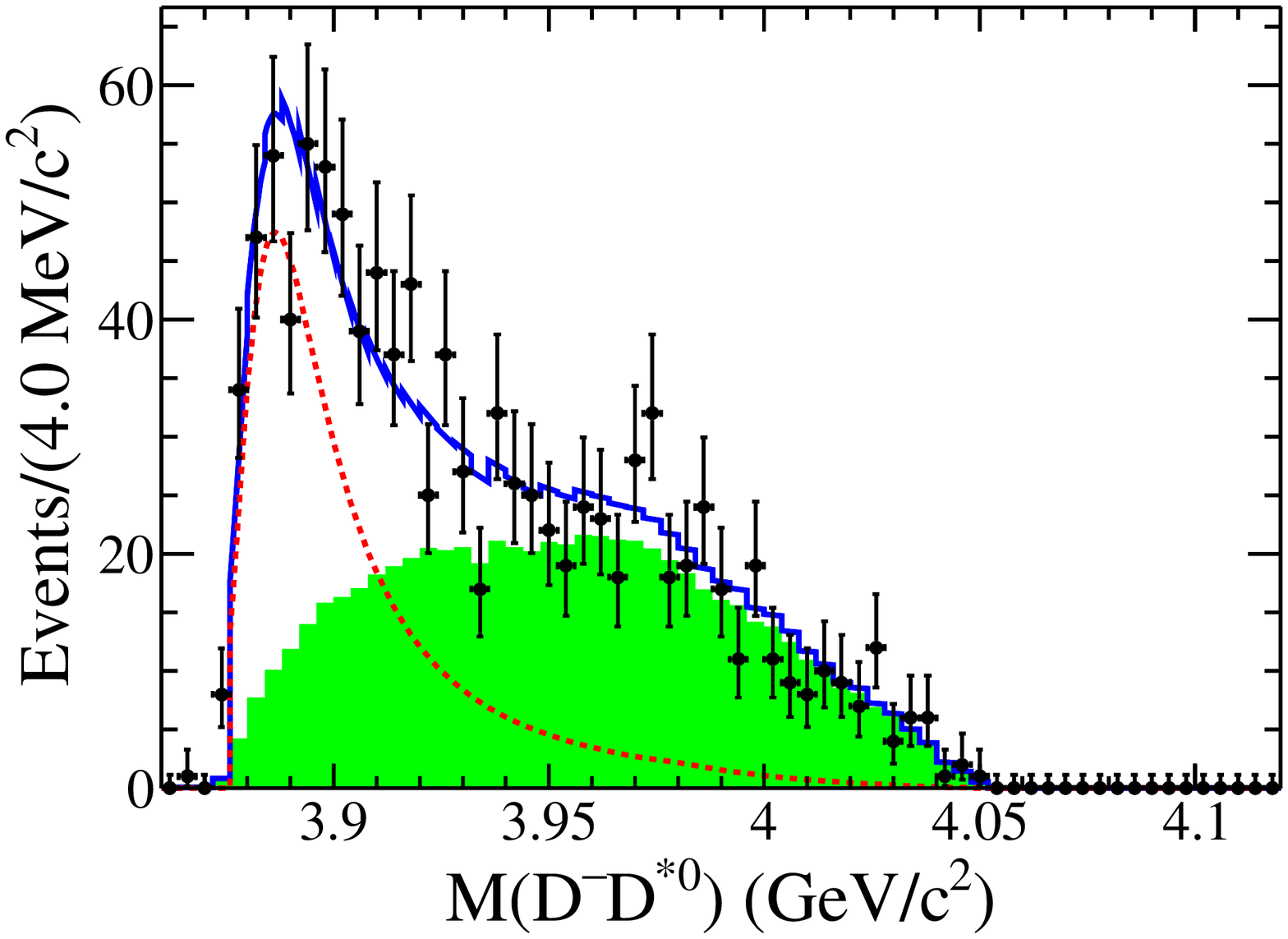}
\put(-50,120){\bf\boldmath (b)}
\end{minipage}
\begin{minipage}[b]{0.45\textwidth}
\epsfig{width=0.98\textwidth,clip=true,file=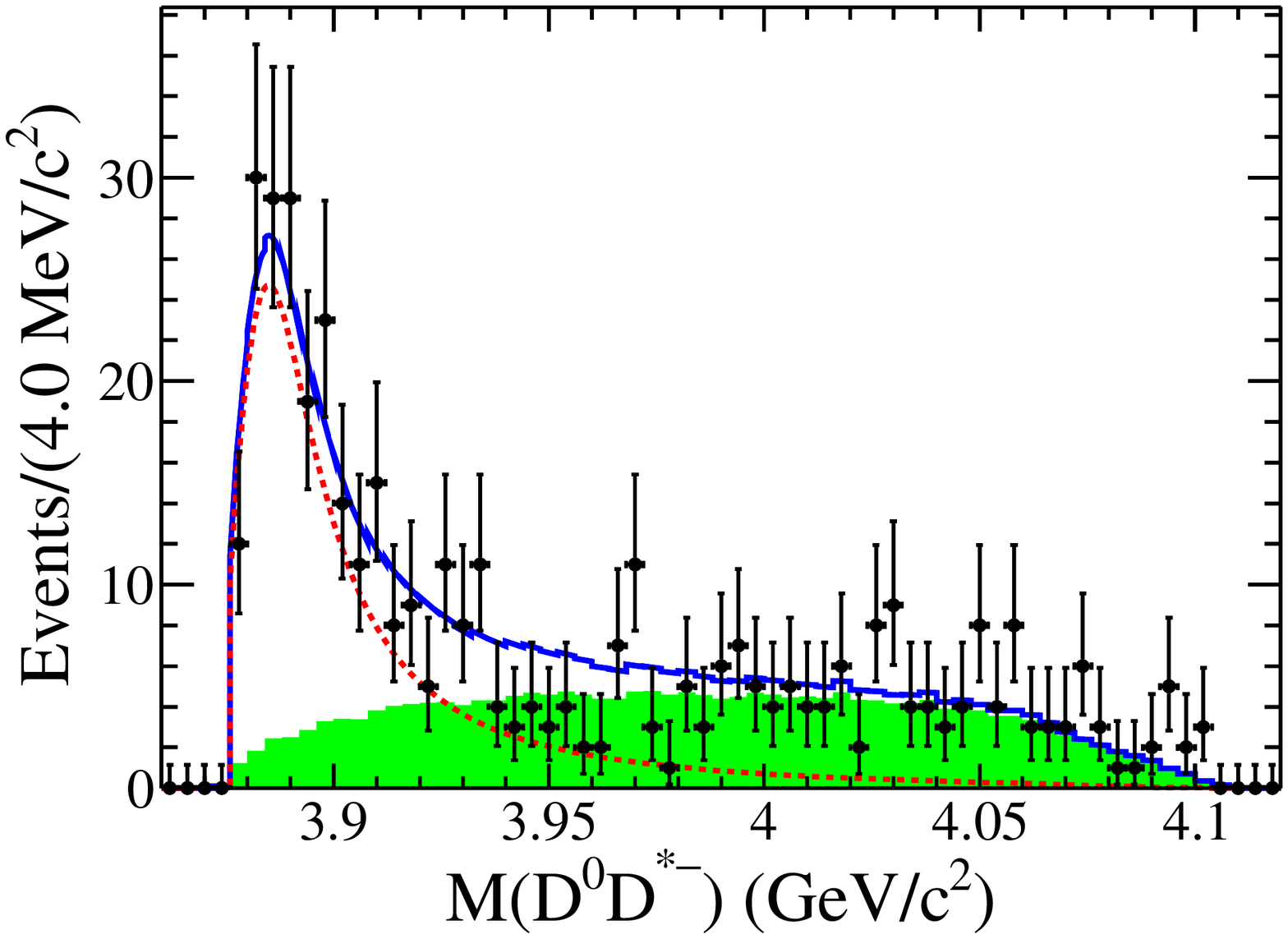}
\put(-50,120){\bf\boldmath (c)}
\end{minipage}
\begin{minipage}[b]{0.45\textwidth}
\epsfig{width=0.98\textwidth,clip=true,file=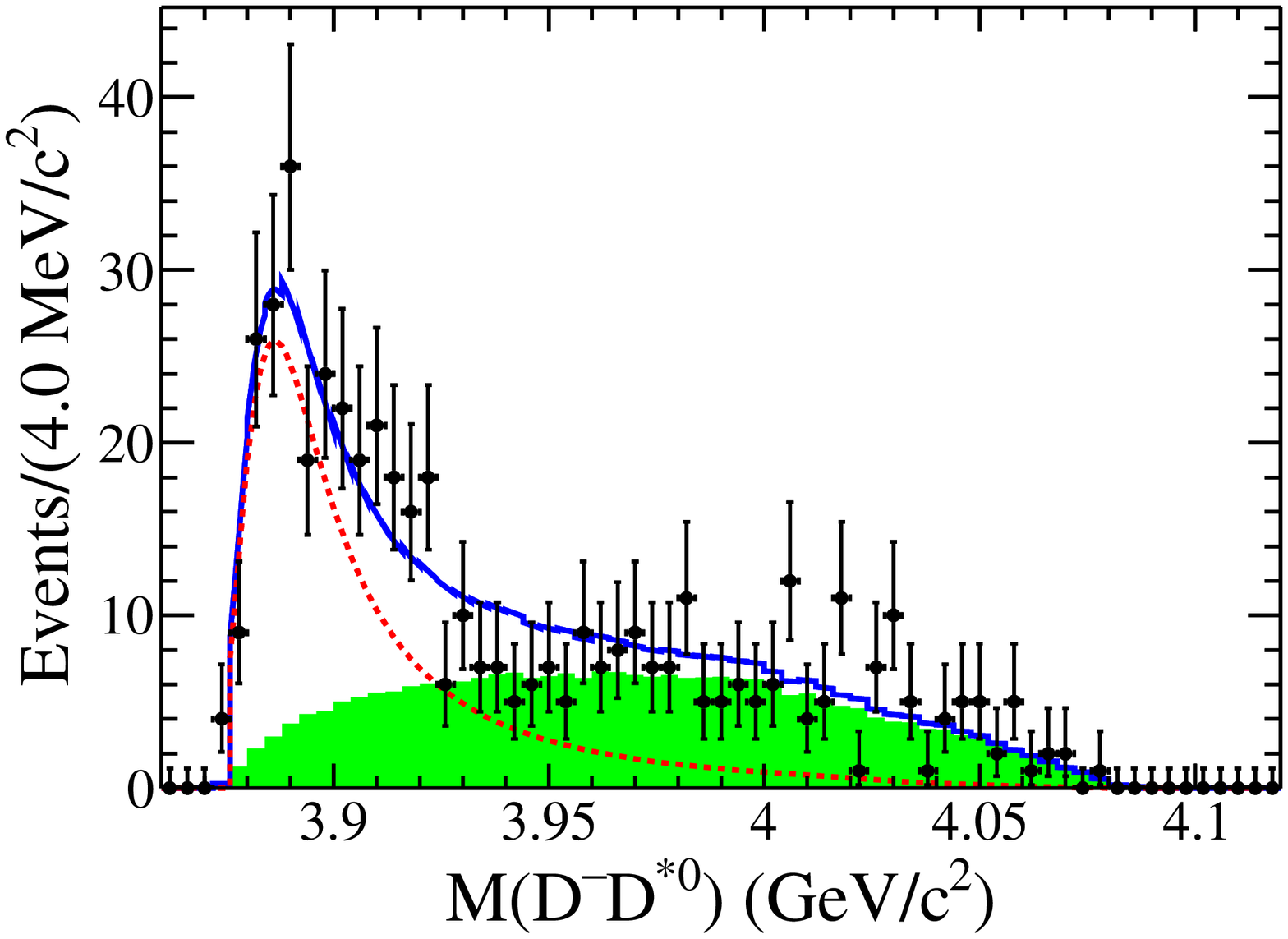}
\put(-50,120){\bf\boldmath (d)}
\end{minipage}
\caption{Simultaneous fits to the $M(D\bar{D}^*)$ distributions of ((a) and (c))
$\pi^+D^0\bar{D}^0$-tagged and ((b) and (d)) $\pi^+D^-D^0$-tagged processes
for ((a) and (b)) data at $\sqrt{s}$=4.23~GeV
and for ((c) and (d)) data at $\sqrt{s}$=4.26~GeV.
The dots with error bars are data
and the lines show the projection of the simultaneous fit to the data.
The solid lines (blue) describe the total fits,
the dashed lines (red) describe the signal shapes and
the green areas describe the background shapes.}
\label{sig_fit}
\end{figure*}

To select the $\pi D\bar{D}^*$ events, we require that $|M_\mathrm{recoil}(D\pi)-M_\mathrm{PDG}(D^{*})| < 30$~MeV/$c^2$.
After imposing all of the above requirements, a peak around 3890~MeV/$c^2$ is clearly visible in
the kinematically constrained $D\bar{D}^*$ mass ($m_{D\bar{D}^*}$) distributions for selected events,
as shown in Fig.~\ref{sig_fit}.
For the $\pi^+D^-D^0$-tagged process, some events from the isospin partner decay channel
$e^+e^-\to\pi^+D^0D^{*-}$ ($D^{*-} \to D^-\pi^0$) can satisfy the above requirements, but
with different reconstruction efficiency and mass resolution.
We treat these as signal events and combine them with the $\pi^+D^-D^0$-tagged process.
For the data sample at $\sqrt{s}$=4.23~GeV, we employ the same event selection criteria and
obtain similar results.

We use the generic MC sample to investigate possible backgrounds.
There is no similar peak found near 3.9~GeV/$c^2$ and the selected events
predominantly have the same final states as $\pi^+(D\bar{D}^*)^-$.
From a study of the Monte Carlo samples of highly excited $D$ states,
we conclude that only the process $e^+e^-\to D_1(2420)\bar{D}, D_1(2420)\to \pi D^*$ can produce a peak near the threshold
in the $D\bar{D}^*$ mass distribution,
although the probability of this is small due to the kinematic boundary.
To examine this possibility, the events are separated into two samples
according to $|\cos\theta_{\pi D}|< 0.5$ and $|\cos\theta_{\pi D}| > 0.5$, where $\theta_{\pi D}$
is the angle between the directions of the bachelor $\pi^+$ and the $D$ meson in the $D\bar{D}^*$ rest frame.
Defining the asymmetry $\mathcal{A} = (n_{>0.5}-n_{<0.5})/(n_{>0.5}+n_{<0.5})$, where $n_{>0.5}$ and $n_{<0.5}$
are the numbers of events in each sample,
we found that the asymmetry in data, $\mathcal{A}_{data}$=0.11$\pm$0.07,
is compatible with the asymmetry expected in signal MC, $\mathcal{A}^{\pi Z_c}_{MC}$ =0.01$\pm$0.01,
and incompatible with the expectations for $D\bar{D}_1(2420)$ MC, $\mathcal{A}^{D\bar{D}_1}_{MC}$ =0.43$\pm$0.01.
Considering the kinematic boundary of this process,
we conclude that the $D\bar{D}_1(2420)$ contribution to our observed Born cross section is smaller than its
relative systematic uncertainty. This is consistent with the ST analysis~\cite{PRL112-022001}.

%% file: signal.tex
\section{Signal Extraction}
\label{sec:signal}
\vspace{-0.4cm}
To extract the resonance parameters and yield of $Z_c(3885)^-$ in the $(D\bar{D}^*)^-$ mass spectrum,
both processes are fitted simultaneously with an unbinned maximum likelihood method
using two different data samples at $\sqrt{s}$=4.23~GeV and $\sqrt{s}$=4.26~GeV.
The $(D\bar{D}^*)^-$ invariant mass distribution is described as
the sum of two probability density functions (PDFs) representing the signal and background.
The signal PDF is given by
\begin{eqnarray}
\begin{aligned}
\mbox{PDF}(m_{D\bar{D}^*}) = \frac{[S(m_{D\bar{D}^*})\otimes R]\epsilon(m_{D\bar{D}^*})}
{\int[S(m_{D\bar{D}^*})\otimes R]\epsilon(m_{D\bar{D}^*})dm_{D\bar{D}^*}},
\end{aligned}
\end{eqnarray}
where the integral is performed over the fit range of the $(D\bar{D}^*)^-$ mass spectrum,
$S(m_{D\bar{D}^*})\otimes R$ is the signal term convolved with the mass resolution,
and $\epsilon(m_{D\bar{D}^*})$ is the reconstruction efficiency.
The background PDF is parameterized by phase space MC simulation.
The signal and background yields and the mass and width of $Z_c(3885)^-$ are determined in the fit.
The mass and width of $Z_c(3885)^-$ are constrained to be the same for both processes.
\subsection{Signal Term}
\vspace{-0.4cm}
The process $e^+e^-\to\pi^{+}Z_c(3885)^{-}$ with $Z_c(3885)^{-}\to \rm{I}$
is described with phase space generalized
for the angular momentum $L$ of the $\pi^{+}-Z_c(3885)^{-}$ system,
where I denotes $D^-D^{*0}$ (labeled as $a$) and $D^0D^{*-}$ (labeled as $b$).
The $Z_c(3885)^{-}$ is described by a mass dependent width Breit-Wigner (MDBW)
parameterization~\cite{PRL96-102002}.
\begin{eqnarray}
\begin{aligned}
&S_{\rm{I}}(m_{D\bar{D}^*})\propto dN/dm_{D\bar{D}^*}\\
&\quad\propto (\kappa^*)^{2L+1}f^2_{L}(\kappa^*)|BW_{\rm{I}}(m_{D\bar{D}^*})|^2,
\end{aligned}
\label{sigpdf}
\end{eqnarray}
where $\kappa^*$ is the momentum of $Z_c(3885)^-$ in the $e^+e^-$ rest frame,
$f_{L}(\kappa^*)$ is the Blatt-Weisskopf barrier factor~\cite{fRi},
\begin{eqnarray}
&BW_{\rm{I}}(m_{D\bar{D}^*})\propto
&\frac{\sqrt{m_{D\bar{D}^*}\Gamma_{\rm{I}}}}
{m^2_{Z_c}-m^2_{D\bar{D}^*}-i\frac{1}{2}m_{Z_c}(\Gamma_{a}+\Gamma_{b})},
\label{RBW}
\end{eqnarray}
$\Gamma_{\rm{I}} = \Gamma_{Z_c}[q_{\rm{I}}^*/q_{\rm{I}}^0]^{2\ell+1}[m_{Z_c}/m_{D\bar{D}^*}][f_{\ell}(q_{\rm{I}}^*)/f_{\ell}(q_{\rm{I}}^0)]^2$,
$q_{\rm{I}}^*$ is the $D$ momentum in the $Z_c(3885)^-$ rest frame,
$\ell$ is the angular momentum of the $(DD^{*})^-$ system, and $q_{\rm{I}}^0 \equiv q_{\rm{I}}^*(m_{Z_c})$.
In the fit, $m_{Z_c}$ and $\Gamma_{Z_c}$ are free parameters,
while $L =0$ and $\ell = 0$ are fixed according to the analysis of angular distributions below.
Parameters of the resolution and efficiency functions, obtained from MC and described below, are fixed in the fit.

\subsection{Reconstruction Efficiency and Mass Resolution}
\begin{figure*}[htbp]
\centering
\begin{minipage}[b]{0.45\textwidth}
\epsfig{width=0.98\textwidth,clip=true,file=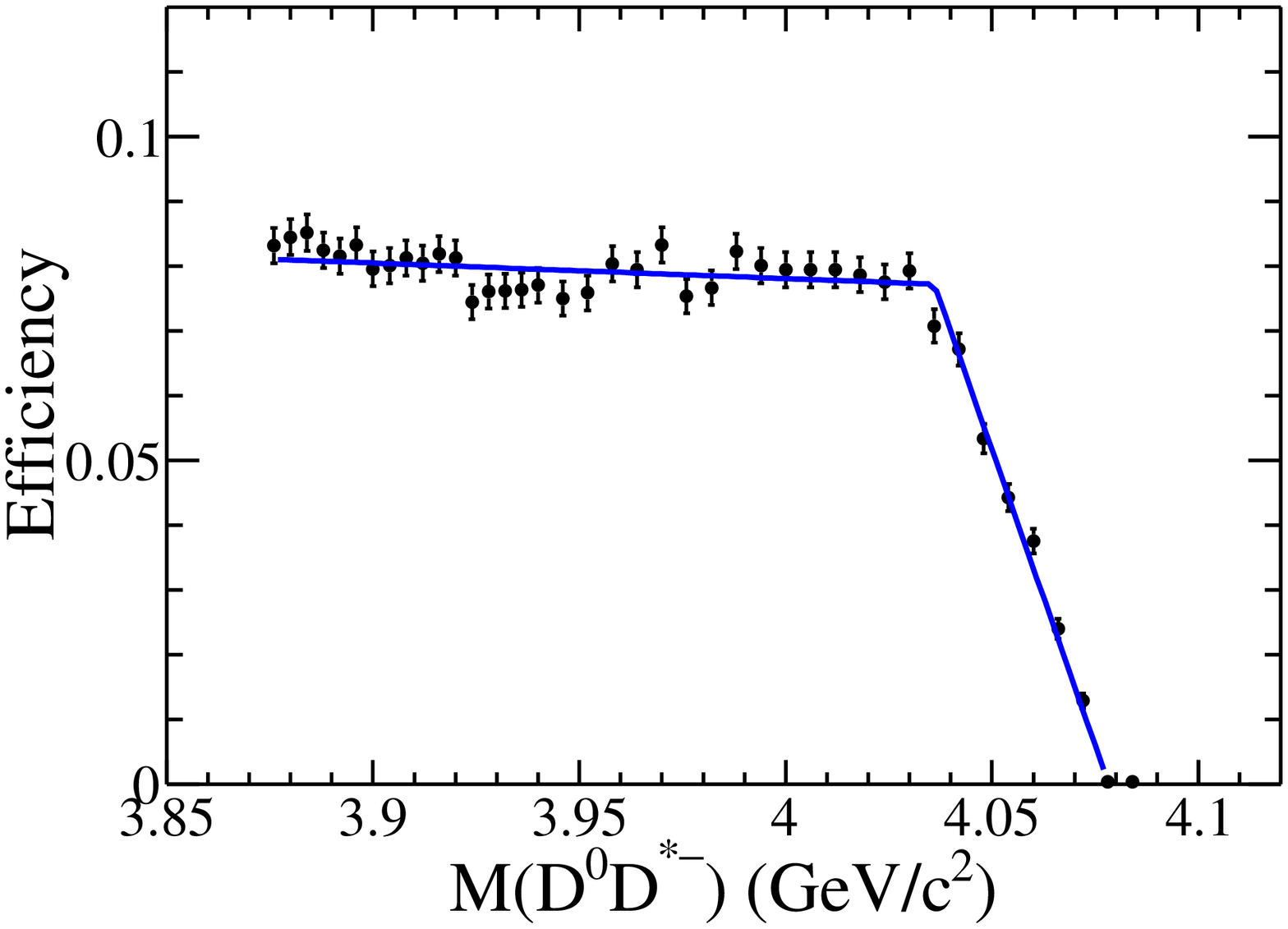}
\put(-50,120){\bf\boldmath (a)}
\end{minipage}
\begin{minipage}[b]{0.45\textwidth}
\epsfig{width=0.98\textwidth,clip=true,file=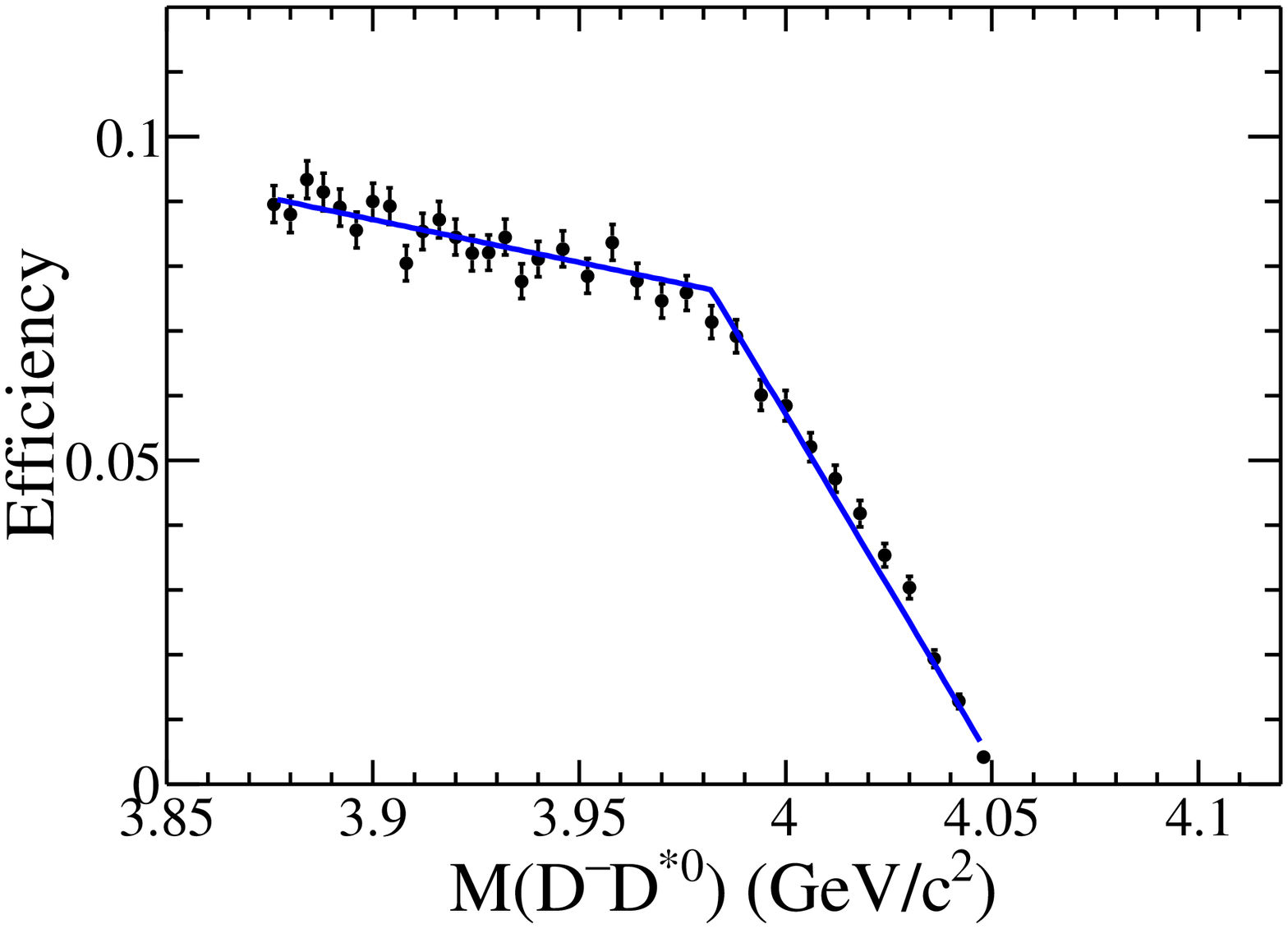}
\put(-50,120){\bf\boldmath (b)}
\end{minipage}
\begin{minipage}[b]{0.45\textwidth}
\epsfig{width=0.98\textwidth,clip=true,file=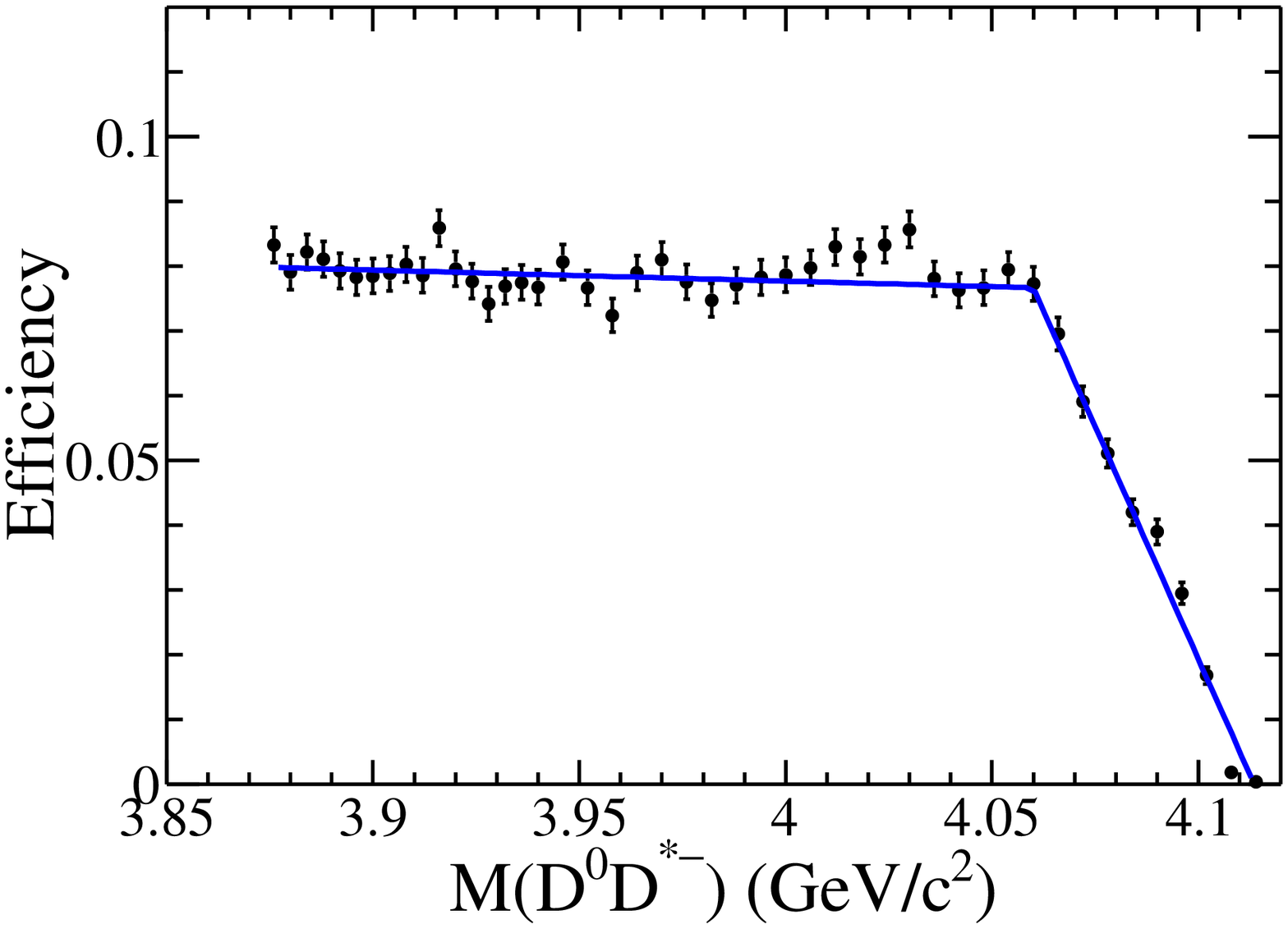}
\put(-50,120){\bf\boldmath (c)}
\end{minipage}
\begin{minipage}[b]{0.45\textwidth}
\epsfig{width=0.98\textwidth,clip=true,file=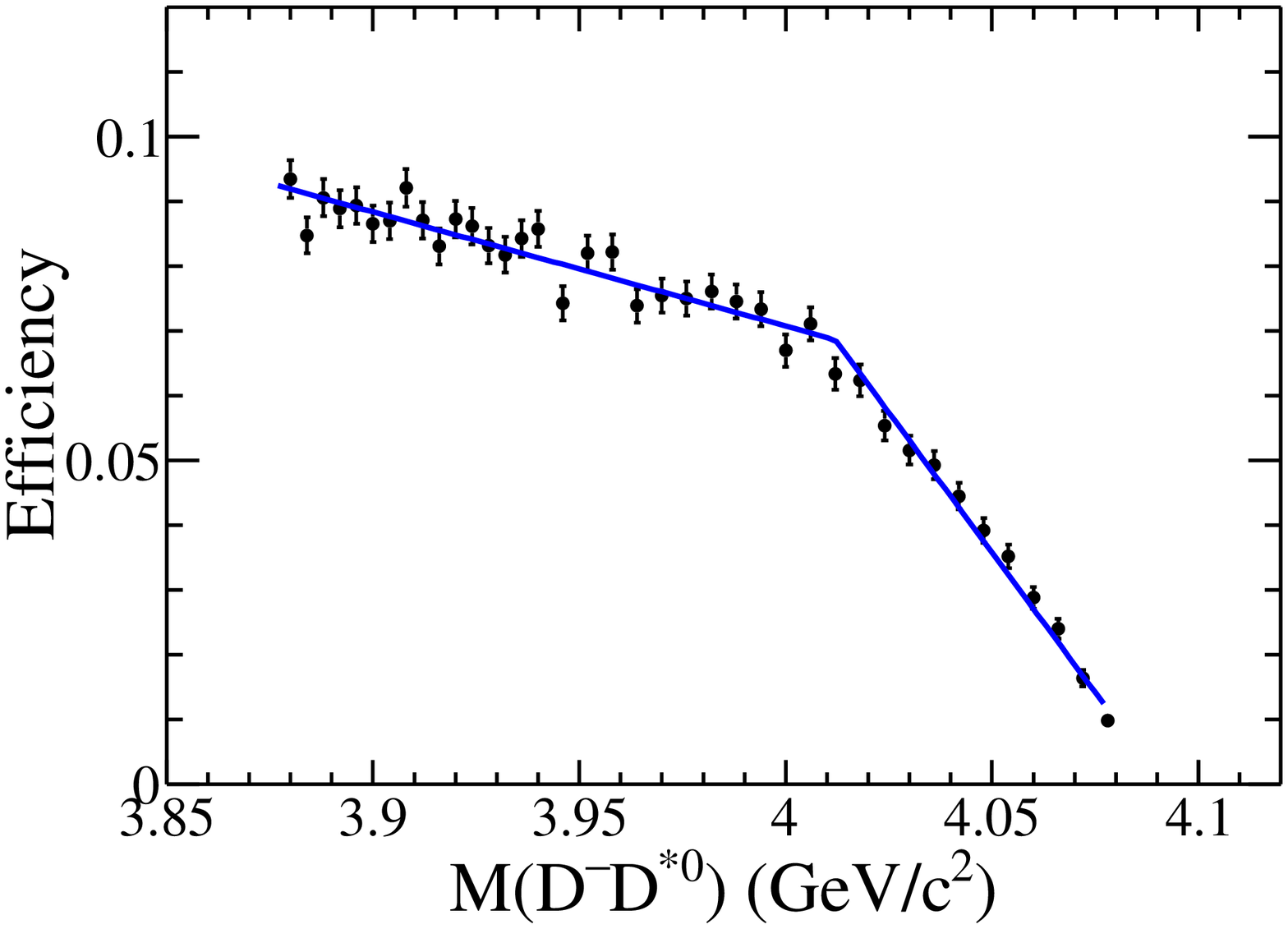}
\put(-50,120){\bf\boldmath (d)}
\end{minipage}
\caption{Distributions of the efficiency versus $M(D\bar{D}^*)$
for ((a) and (c)) $\pi^+D^0\bar{D}^0$-tagged and ((b) and (d)) $\pi^+D^-D^0$-tagged processes
at ((a) and (b)) $\sqrt{s}$=4.23~GeV and ((c) and (d)) $\sqrt{s}$=4.26~GeV.
The dots with error bars are the efficiencies determined from MC.
The curves show the fits with a piecewise linear function.}
\label{eff_curve}
\end{figure*}

\vspace{-0.4cm}
In order to obtain the reconstruction efficiency and mass resolution,
we generate a set of MC samples for $e^+e^-\to\pi^+Z^-_c~(Z^-_c\to(D\bar{D}^*)^-)$,
each with a fixed mass value, zero width and $J^P=1^+$ of the $Z^-_c$,
and subject these MC samples to the same event selection criteria.
The isospin channel $e^+e^-\to\pi^+D^0D^{*-}$ ($D^{*-} \to D^-\pi^0$) can feed into the $\pi^+D^-D^0$-tagged process.
We therefore generate two corresponding MC samples by assuming the same decay branching fraction between the process
$Z_c^-\to D^-D^{*0}$ and $Z_c^-\to D^0D^{*-}$.
The reconstruction efficiency is estimated using the sum of the two MC samples, as shown in Fig.~\ref{eff_curve}.

\begin{figure*}[hbtp]
\centering
\begin{minipage}[b]{0.45\textwidth}
\epsfig{width=0.98\textwidth,clip=true,file=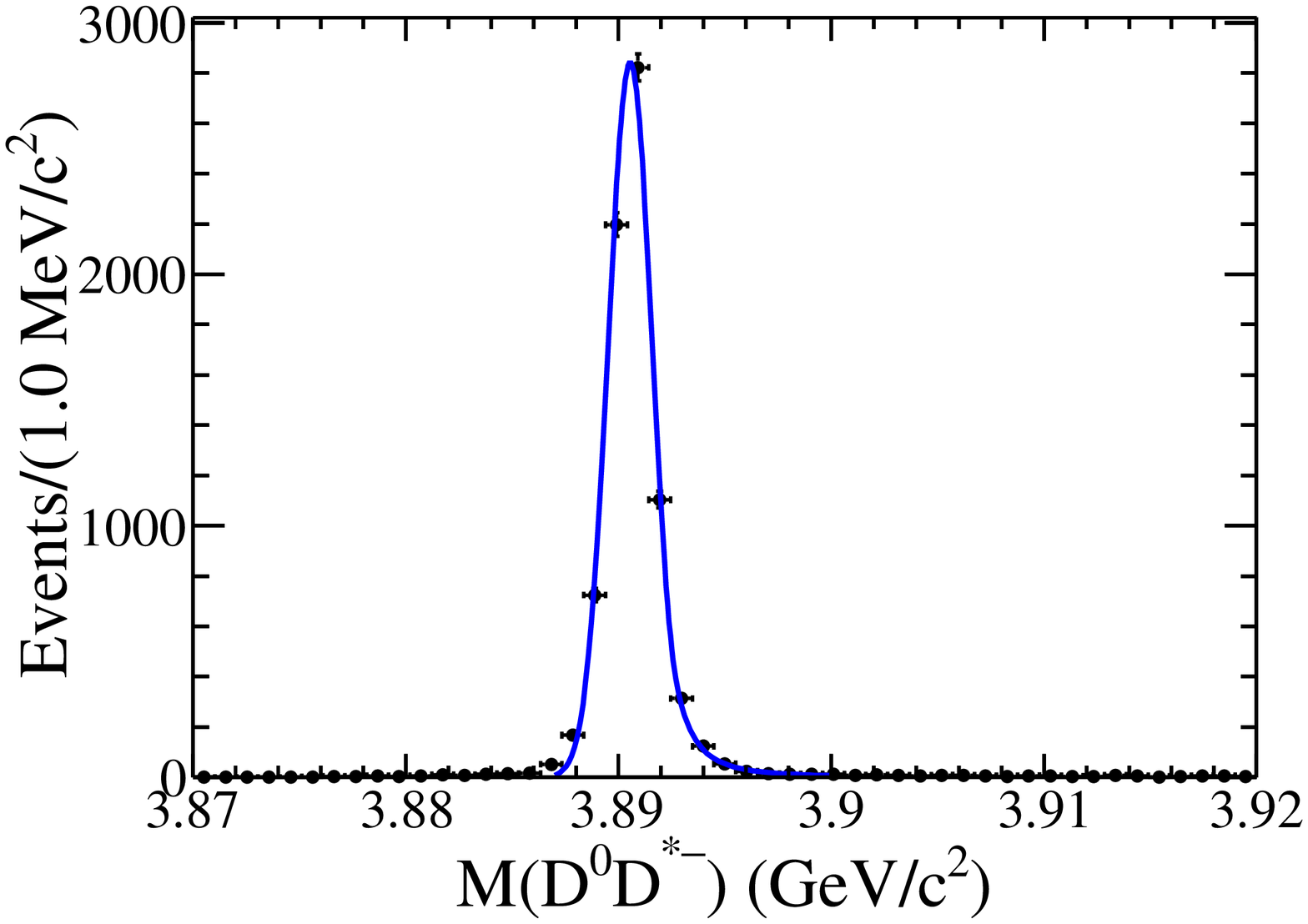}
\put(-50,120){\bf\boldmath (a)}
\end{minipage}
\begin{minipage}[b]{0.45\textwidth}
\epsfig{width=0.98\textwidth,clip=true,file=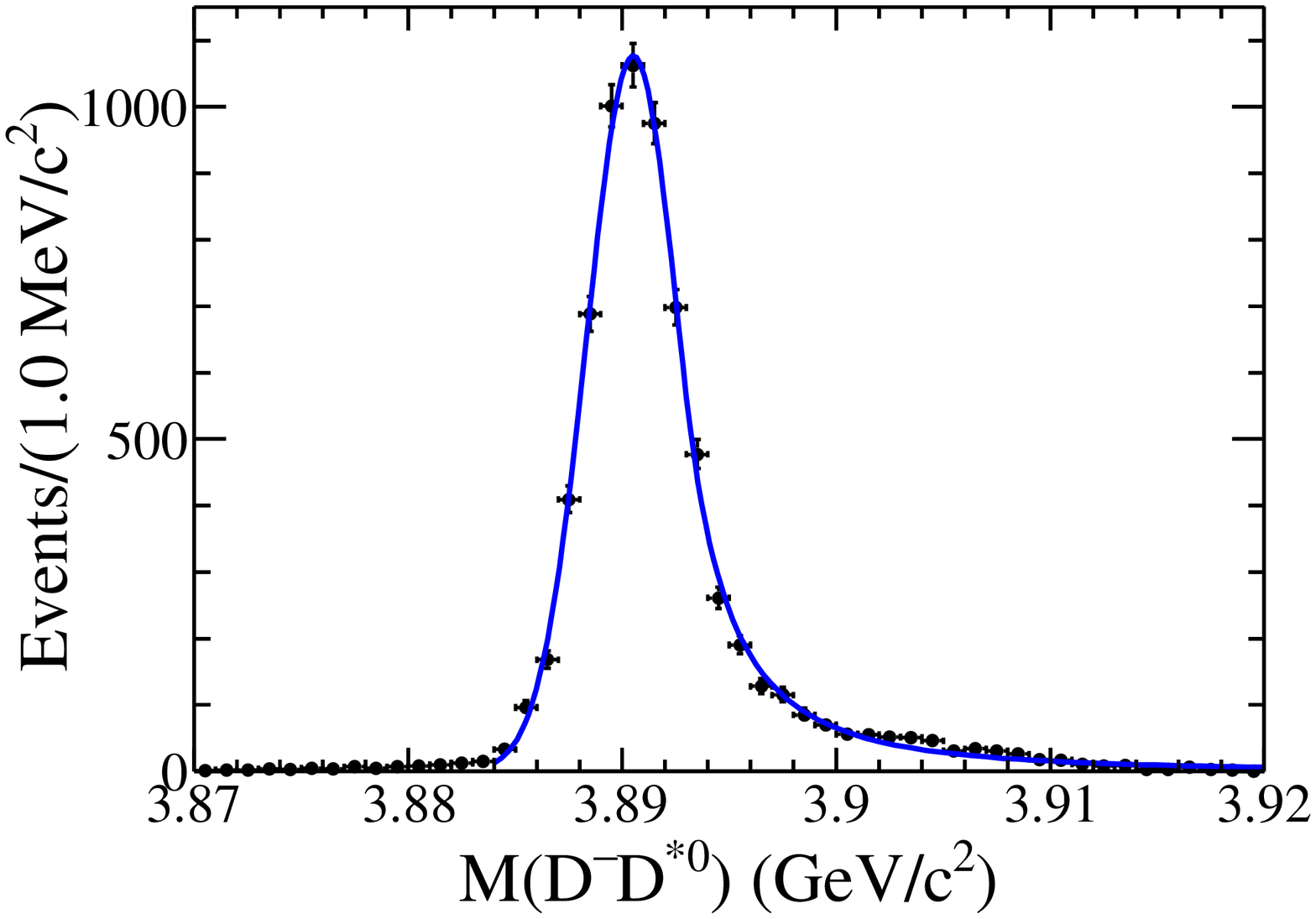}
\put(-50,120){\bf\boldmath (b)}
\end{minipage}
\begin{minipage}[b]{0.45\textwidth}
\epsfig{width=0.98\textwidth,clip=true,file=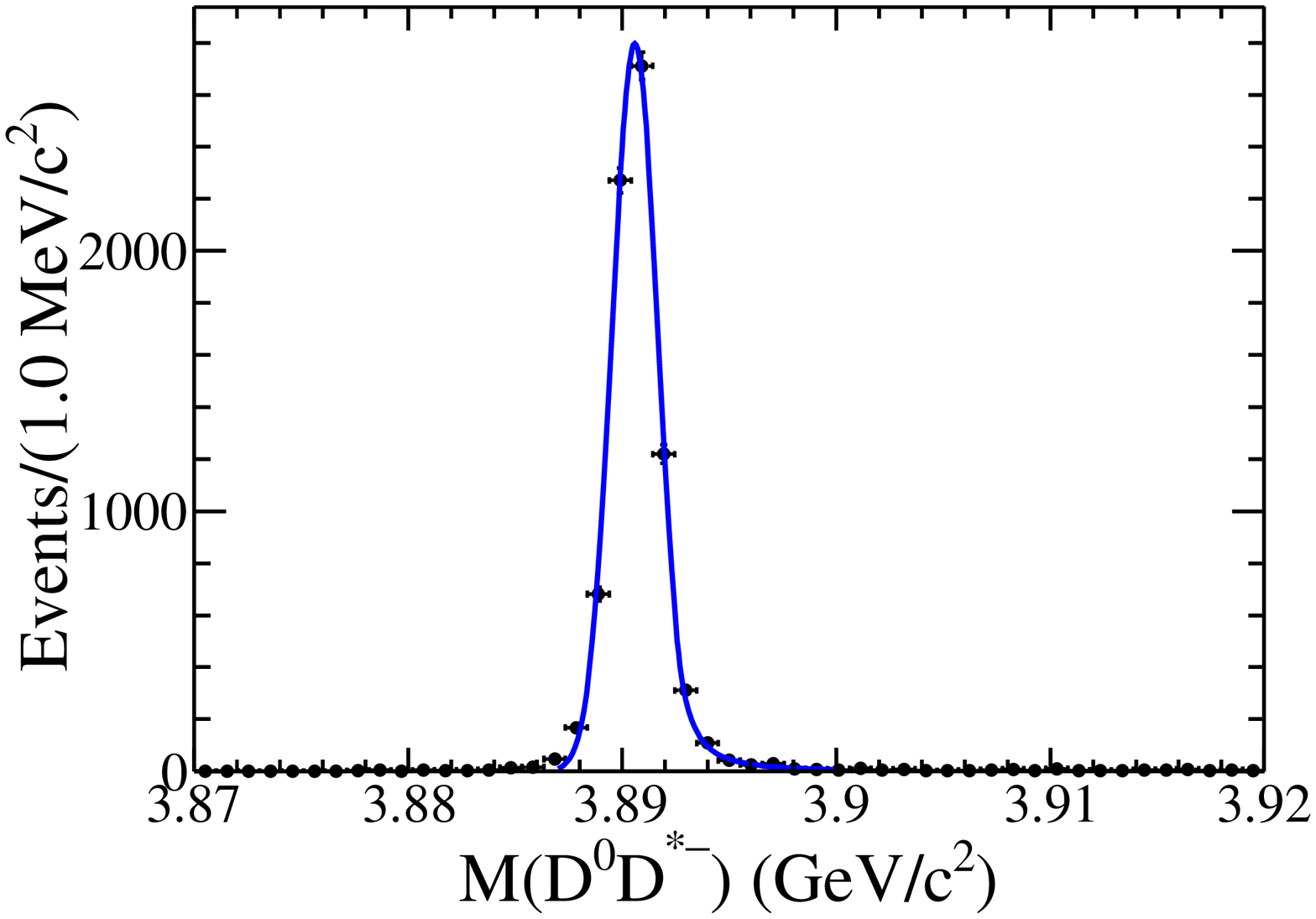}
\put(-50,120){\bf\boldmath (c)}
\end{minipage}
\begin{minipage}[b]{0.45\textwidth}
\epsfig{width=0.98\textwidth,clip=true,file=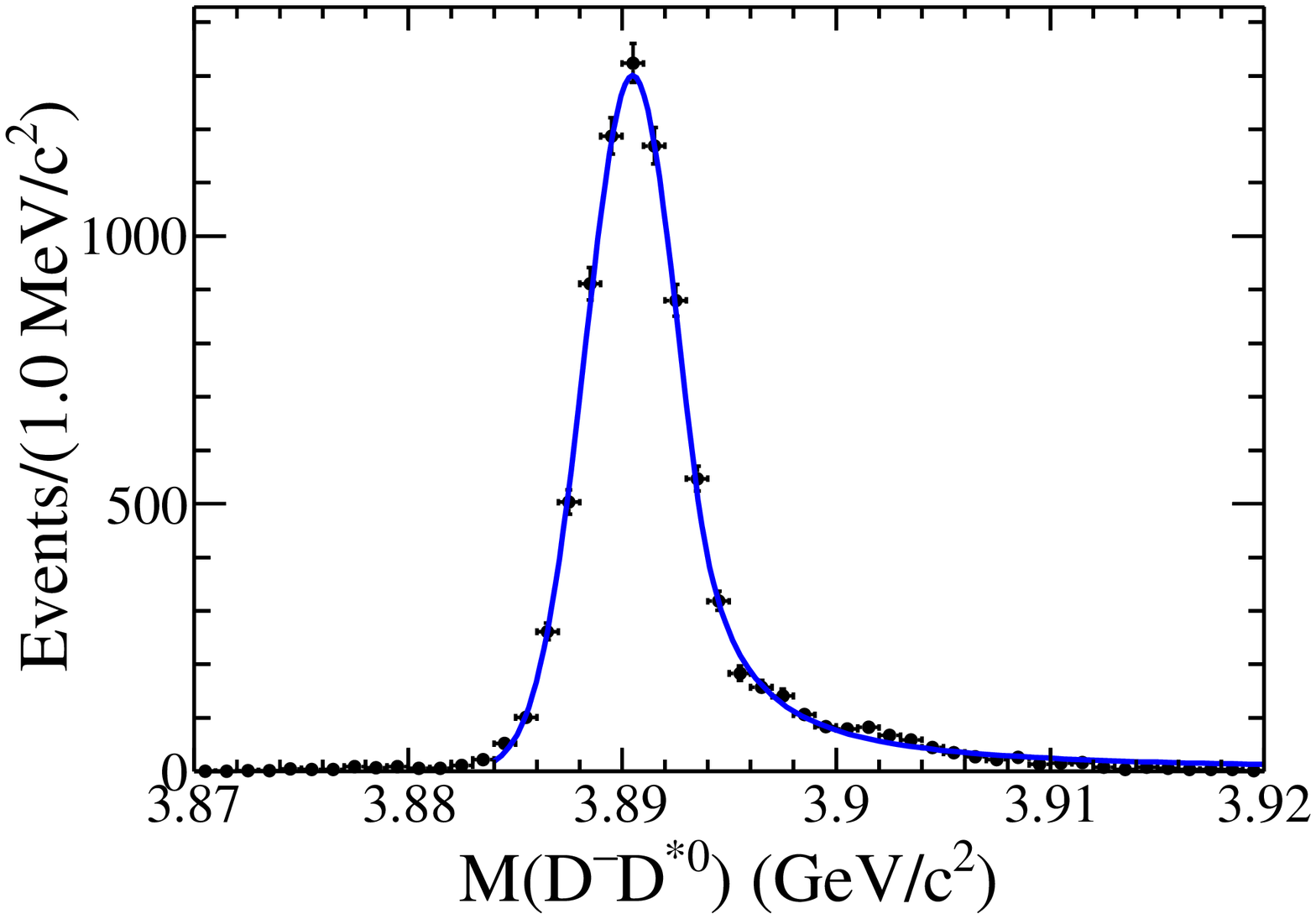}
\put(-50,120){\bf\boldmath (d)}
\end{minipage}
\caption{Fits to the mass resolution at 3890~MeV for
((a) and (c)) $\pi^+D^0\bar{D}^0$-tagged and ((b) and (d)) $\pi^+D^-D^0$-tagged processes
at ((a) and (b)) $\sqrt{s}$=4.23~GeV and ((c) and (d)) $\sqrt{s}$=4.26~GeV.
The dots with error bars show the distributions of mass resolutions obtained from MC,
the curves show the fits.}
\label{resolution}
\end{figure*}

MC samples for $e^+e^-\to\pi^+Z^-_c~(Z^-_c\to(D\bar{D}^*)^-)$ are used to determine the mass resolution.
The mass and width of $Z_c$ are set to be 3890~MeV/$c^2$ and 0~MeV, respectively.
The mass resolution for the $\pi^+D^0\bar{D}^0$-tagged process is described
by a Crystal Ball (CB) function~\cite{CB}.
Since the $\pi^+D^-D^0$-tagged process contains two isospin processes,
the mass resolution is represented by a sum of two CB functions with a common mean and different widths.
The fit results for both processes are shown in Fig.~\ref{resolution}.
The resolution for the $\pi^+D^0\bar{D}^0$-tagged process is determined by
the fit to be 1.1$\pm$0.1~MeV/$c^2$,
while the resolution for the $\pi^+D^-D^0$-tagged process is
calculated to be 2.2$\pm$0.1~MeV/$c^2$ using the equation $f_1 \sigma_1 + (1-f_1)\sigma_2$,
where $\sigma_1$ and $\sigma_2$ are the individual widths of each of the two CB functions and
$f_1$ is the fractional area of the first CB function.

\subsection{Fit Results }
\vspace{-0.4cm}
As shown in Fig.~\ref{sig_fit}, we perform a simultaneous fit to the $M(D\bar{D}^*)$ distributions
for the $\pi^+D^0\bar{D}^0$-tagged and $\pi^+D^-D^0$-tagged processes
with $\sqrt{s}$=4.23~GeV and $\sqrt{s}$=4.26~GeV data samples.
The statistical significance of $Z_c(3885)^-$, estimated by the difference of log-likelihood
values with and without signal terms in the fit, is greater than 10$\sigma$.
The mass and width of $Z_c(3885)^-$ are fitted to be
$M_{Z_c(3885)}$ = (3890.3$\pm$0.8)~MeV/$c^2$ and $\Gamma_{Z_c(3885)}$ = (31.5$\pm$3.3)~MeV,
where the errors are statistical only.
Since the resulting mass and width might be different from the actual resonance properties
due to the parameterization function of $Z_c(3885)$,
we calculate the pole position ($P = M_{\rm{pole}} -i\Gamma_{\rm{pole}}/2$) of $Z_c(3885)$
which is the complex number where the denominator of $BW_{\rm{I}}(m_{D\bar{D}^*})$ is zero,
and regard $M_{\rm{pole}}$ and $\Gamma_{\rm{pole}}$ as the final result.
The corresponding pole mass ($M_{\rm{pole}}$) and width ($\Gamma_{\rm{pole}}$) of
$Z_c(3885)$ are $M_{\rm{pole}}$ = (3881.7$\pm$1.6)~MeV/$c^2$ and $\Gamma_{\rm{pole}}$ = (26.6$\pm$2.0)~MeV, respectively.

\subsection{Angular Distribution}
\vspace{-0.4cm}
The quantum number $J^P$ assignment for $Z_c(3885)^-$ is investigated by examining the distribution of \textbar \rm{cos}$\theta_{\pi}$\textbar,
where $\theta_\pi$ is the $\pi^+$ polar angle relative to the beam direction in the center-of-mass frame.
If $J^P=1^+$, the relative orbital angular momentum of the $\pi^+$-$Z_c(3885)^{-}$ system could be either $S$-wave or $D$-wave.
If we neglect the small contribution of $D$-wave due to the closeness of the threshold,
the \textbar \rm{cos}$\theta_{\pi}$\textbar ~distribution is expected to be flat.
If $J^P=0^-$ ($1^-$), the $\pi^{+}$-$Z_c(3885)^{-}$ system occurs via a $P$-wave
and the \textbar \rm{cos}$\theta_{\pi}$\textbar ~is expected to follow $\rm{sin}^2\theta_{\pi}$ (1+$\rm{cos}^2\theta_{\pi}$) distribution.

\begin{figure*}[htbp]
\centering
\begin{minipage}[b]{0.48\textwidth}
\epsfig{width=0.98\textwidth,clip=true,file=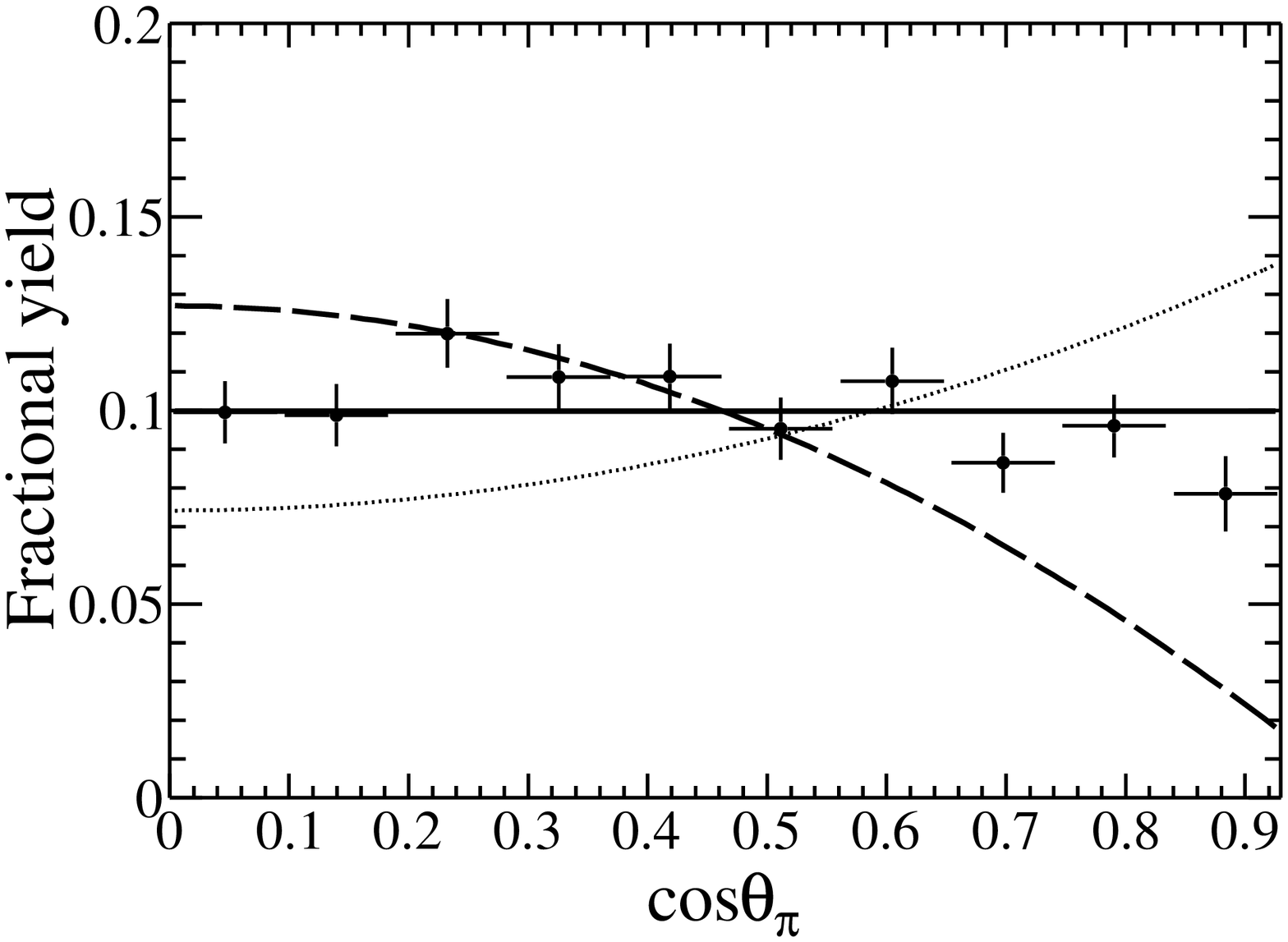}
\put(-50,120){\bf\boldmath (a)}
\end{minipage}
\begin{minipage}[b]{0.48\textwidth}
\epsfig{width=0.98\textwidth,clip=true,file=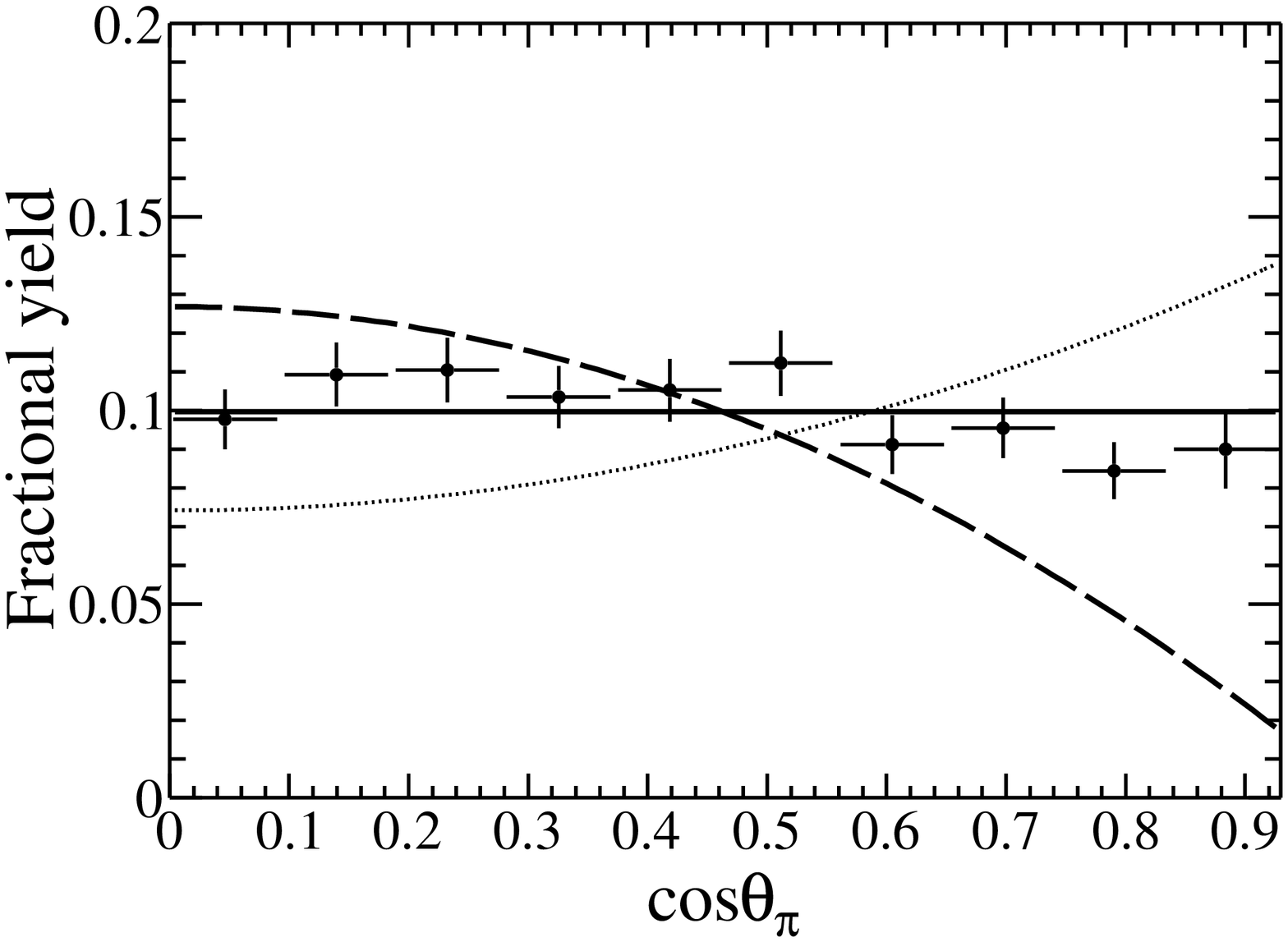}
\put(-50,120){\bf\boldmath (b)}
\end{minipage}
\caption{Fits to \textbar \rm{cos}$\theta_{\pi}$\textbar\ distributions for (a) $\pi^+D^0\bar{D}^0$-tagged
and (b) $\pi^+D^-D^0$-tagged processes.
The dots with error bars show the combined data corrected for detection efficiency at $\sqrt{s}$=4.23~GeV and $\sqrt{s}$=4.26~GeV,
the solid lines show the fits using $J^P=1^+$ hypothesis,
and the dashed and dotted curves are for the fits with $J^P=0^-$ and $J^P=1^-$ hypothesis, respectively.}
\label{Eff_cospi}
\end{figure*}

The \textbar \rm{cos}$\theta_{\pi}$\textbar ~distribution of data is plotted with
the efficiency corrected signal yield of combined data samples at $\sqrt{s}$=4.23~GeV and $\sqrt{s}$=4.26~GeV
in ten \textbar \rm{cos}$\theta_{\pi}$\textbar ~bins,
where the signal yields in different bin are extracted with the same simultaneous fit method described above.
Figures~\ref{Eff_cospi} (a) and (b) show the \textbar \rm{cos}$\theta_{\pi}$\textbar ~distribution for
$\pi^+D^0\bar{D}^0$-tagged process and $\pi^+D^-D^0$-tagged process, respectively.
The data agrees well with the flat distribution expected
for $J^P=1^+$ ($\chi^2$/NDF = 16.5/9 for the $\pi^+D^0\bar{D}^0$-tagged process
and 12.8/9 for the $\pi^+D^-D^0$-tagged process) and disagrees with the $\rm{sin}^2\theta_{\pi}$
distribution expected for $J^P=0^-$ ($\chi^2$/NDF = 103.1/9 for the $\pi^+D^0\bar{D}^0$-tagged process
and 104.9/9 for the $\pi^+D^-D^0$-tagged process) and $J^P=1^-$
($\chi^2$/NDF = 106.3/9 for the $\pi^+D^0\bar{D}^0$-tagged process
and 104.9/9 for the $\pi^+D^-D^0$-tagged process), where NDF is the number of degrees of freedom in the fit.

\subsection{Born Cross Section}
\vspace{-0.4cm}
For the $\pi^+D^0\bar{D}^0$-tagged process, the Born cross section
times the $(D\bar{D}^*)^-$ branching fraction of $Z_c(3885)^-$ ($\sigma\times Br$) can be calculated by
\begin{eqnarray}
\begin{aligned}
\sigma(e^+e^-\to\pi^{\pm}Z_c(3885)^{\mp})\times Br(Z_c(3885)^{\mp}\to (D\bar{D}^*)^{\mp})\\
= \frac{N}{\mathcal{L}(1+\delta^{r})(1+\delta^{v})
\sum_{\substack {i,j}}\epsilon_{ij}Br_{i}Br_{j}Br(D^{*-}\to\pi^-\bar{D}^0)I},
\label{eq1}
\end{aligned}
\end{eqnarray}
where $N$ is the signal yield, $\mathcal{L}$ is the integrated luminosity,
$\epsilon_{ij}$ is the signal efficiency for the $\pi^+D^0\bar{D}^0$-tagged process listed in Table~\ref{tab:eff_D0D0}
of Appendix~\ref{sec:appA},
where the subscripts $i, j = 0 \ldots 3$ denote the neutral $D$ final state,
$Br_{i}$ is the individual branching fraction for $D$ decay from PDG~\cite{pdg},
the radiative correction factor ($1+\delta^{r}$) is determined by
the measurement of the line shape of $\sigma(e^+e^-\to\pi D\bar{D}^*)$~\cite{PRL112-022001},
the vacuum polarization factor ($1+\delta^{v}$) is considered in the MC simulation~\cite{EPJC-66-585}
and $I = Br(Z_c(3885)^-\to D^0D^{*-})/Br(Z_c(3885)^-\to (D\bar{D}^*)^-)$ = 0.5, assuming isospin symmetry.
The value of all above variables are listed in Table~\ref{tab:cros}.

Since the $\pi^+D^-D^0$-tagged process contains two processes of
$Z_c(3885)^-\to D^-D^{*0}$ with $D^{*0}\to\pi^0D^0$ (labeled as $\alpha$)
and $Z_c(3885)^-\to D^0D^{*-}$ with $D^{*-}\to\pi^0D^-$ (labeled as $\beta$),
the Born cross section times the $(D\bar{D}^*)^-$ branching fraction of $Z_c(3885)^-$ can be given by
\begin{widetext}
\begin{eqnarray}
\begin{aligned}
&\sigma(e^+e^-\to\pi^{\pm}Z_c(3885)^{\mp})\times Br(Z_c(3885)^{\mp}\to (D\bar{D}^*)^{\mp})\\
&= \frac{N}{\mathcal{L}(1+\delta^{r})(1+\delta^{v})
(\sum_{\substack {i,j}}\epsilon^{\alpha}_{ij}Br_{i}Br_{j}Br(D^{*0}\to\pi^0D^0)
+\sum_{\substack {i,j}}\epsilon^{\beta}_{ij}Br_{i}Br_{j}Br(D^{*-}\to\pi^0D^-))I},\\
\end{aligned}
\end{eqnarray}
\end{widetext}
where $\epsilon^{\alpha}_{ij}$ and $\epsilon^{\beta}_{ij}$ are the signal efficiency for the two $\pi^+D^-D^0$-tagged processes,
listed in Table~\ref{tab:eff_D0Dma} and~\ref{tab:eff_D0Dmb} of Appendix~\ref{sec:appA},
the subscripts $i$ and $j$ denote the $D^-$ and $D^0$ final states,
respectively, with $i = A \ldots F$ and $j = 0 \ldots 3$,
$Br(D^{*0}\to\pi^0D^0)$ = (61.9$\pm$2.9)\% and $Br(D^{*-}\to\pi^0D^-)$ = (30.7$\pm$0.5)\%~\cite{pdg}.
The value of all above variables are listed in Table~\ref{tab:cros}.

We also add a $Z_c(4020)^-$ in the fit with mass and width fixed to the BESIII measurement~\cite{PRL111-242001}.
The fit prefers the presence of a $Z_c(4020)^-$ with a statistical significance of 1.0$\sigma$.
We determine the upper limit on $\sigma\times Br$ at the 90\% confidence level (C.L.),
where the probability density function from the fit is smeared by a Gaussian function
with a standard deviation of the relative systematic error in the $\sigma\times Br$ measurement.
We obtain $\sigma(e^+e^-\to\pi^{\pm}Z_c(4020)^{\mp})\times Br(Z_c(4020)^{\mp}\to (DD^*)^{\mp})<$
18~pb at $\sqrt{s}$=4.23~GeV and $<$15~pb at $\sqrt{s}$=4.26~GeV, respectively, at 90\% C.L..

\begin{table*}[htbp]
\begin{center}
\caption{Summary of the product of Born cross sections times the $(D\bar{D}^*)^-$ branching
fraction of $Z_c(3885)^-$ ($\sigma\times Br$), the errors are statistical only.
}
\begin{tabular}{lcccc}
\hline
\hline
\multirow{2}{*}{}
& \multicolumn{2}{c}{$\pi^+D^0\bar{D}^0$-tagged process}
& \multicolumn{2}{c}{$\pi^+D^-D^0$-tagged process}
\\
\cline{2-3}
\cline{4-5}
& 4.23~GeV
& 4.26~GeV
& 4.23~GeV
& 4.26~GeV
\\
\hline
\hline
$N$			      &  384$\pm$30  &   207$\pm$18  &  418$\pm$34  & 239$\pm$22  \\
$\mathcal{L}$ (pb$^{-1}$)     &  1091.7        &   825.7         &  1091.7        & 825.7         \\
1+$\delta^{r}$		      &  0.89 & 0.92     & 0.89 & 0.92       \\
1+$\delta^{v}$		      &  1.056 & 1.054                   &     1.056 & 1.054                   \\
$\sigma\times Br$ (pb)	      &  147.5$\pm$11.5 & 109.2$\pm$9.7 & 136.6$\pm$11.0 & 107.5$\pm$9.7  \\
\hline
\hline
\end{tabular}
\label{tab:cros}
\end{center}
\end{table*}

%% file: syserr.tex
\section{Systematic Uncertainties}
\label{sec:syserr}
\vspace{-0.4cm}
The systematic uncertainties for the pole mass and width of $Z_c(3885)^-$,
and the product of Born cross section times the $(D\bar{D}^*)^-$ branching
fraction of $Z_c(3885)^-$ ($\sigma\times Br$) are
described below and summarized in Table~\ref{table:syserr}.
The total systematic uncertainty is obtained by summing all individual contributions in quadrature.

\noindent\textbf{Beam Energy:} In order to obtain the systematic uncertainty related to the beam energy,
we repeat the whole analysis by varying the beam energy with $\pm1$~MeV in the kinematic fit.
The largest difference on the pole mass, width and the signal yields is taken as a systematic uncertainty. 

\noindent\textbf{Mass Calibration:}
The uncertainty from the mass calibration is estimated with the difference between
the measured and nominal $D^*$ masses.
We fit the $D^*$ mass spectra calculated with the output momentum of the kinematic fit described
in the Sec.~\ref{sec:selection} after removing the D* mass constraint.
The deviation of the resulting $D^*$ mass to the nominal values is found to be 0.84$\pm$0.16~MeV/$c^2$.
The systematic uncertainty due to the mass calibration is taken to be 1.0~MeV/$c^2$.

\noindent$\bm{\mathcal{L}(1+\delta^{r})(1+\delta^{v}):}$
The integrated luminosities of the data samples are measured
using large angle Bhabha events, with an estimated uncertainty of 1.0\%~\cite{arxiv:150303408}.
The systematic uncertainty of the radiative correction factor is estimated by changing
the parameters of the line shape of $\sigma(e^+e^-\to\pi D\bar{D}^*)$ within errors.
We assign 4.6\% as the systematic uncertainty due to the radiative correction factor according to Ref.~\cite{PRL112-022001}.
The systematic uncertainty of the vacuum polarization factor is 0.5\%~\cite{EPJC-66-585}.

\noindent\textbf{Signal shape:}
The systematic uncertainty associated with the $Z_c(3885)^-$ signal shape
is evaluated by repeating the fit on the $M(D\bar{D}^*)$ distribution
with a mass constant width BW line shape (MCBW, $\frac{1}{m^2_{Z_c}-m^2_{D\bar{D}^{*}}-im_{Z_c}\Gamma_{Z_c}}$) for $Z_c(3885)^-$ signal.
The resulting difference to the nominal results are taken as a systematic uncertainty.

\noindent\textbf{$Z_c(4020)^-$ signal} The systematic uncertainty associated with the possible existence of the $Z_c(4020)^-$
in our data is estimated by adding the $Z_c(4020)^-$ in the fit. The difference of fit results is taken
as a systematic uncertainty.

\noindent\textbf{Background shape:} The systematic uncertainty due to the background shape is investigated
by repeating the fit with function
$f_{bkg}(m_{D\bar{D}^*})\propto (m_{D\bar{D}^*} - M_{min})^c(M_{max}- m_{D\bar{D}^*})^d$~\cite{PRL112-022001}
for the background line shape,
where $M_{min}$ and $M_{max}$ are the minimum and maximum kinematically allowed masses, respectively,
$c$ and $d$ are free parameters. The resulting difference to the nominal results is taken as a systematic uncertainty.

\noindent\textbf{Fit bias:}
To assess a possible bias due to the fitting procedure,
we generate 200 fully reconstructed data-size samples
with the parameters set to the values (input values) returned by the fit to data.
Then we fit these samples using the same procedures as we fit the data,
and the resulting distribution of every fitted parameter with a Gaussian function.
The difference between the mean value of Gaussian and the input value is taken as a systematic uncertainty of the fit bias. 

\noindent \textbf{Signal region of DT:} In order to obtain the systematic uncertainty
related to the selection of the signal region of double $D$ tag,
we repeat the whole analysis by changing the signal region in the $M(\bar{D})$ versus $M(D)$ plane from the nominal region
to $-15<\Delta\hat{M}<10$~MeV/$c^2$ ($|\Delta M| < 30$~MeV/$c^2$) and $-25<\Delta\hat{M}<20$~MeV/$c^2$ ($|\Delta M| < 60$~MeV/$c^2$)
for $\pi^+D^0\bar{D}^0$-tagged,
and $-14<\Delta\hat{M}<11$~MeV/$c^2$ ($|\Delta M| < 28$~MeV/$c^2$) and $-20<\Delta\hat{M}<17$~MeV/$c^2$ ($|\Delta M| < 42$~MeV/$c^2$)
for $\pi^+D^-D^0$-tagged processes. 
The largest difference of fit results is taken as a systematic uncertainty.

\noindent \textbf{Efficiency related:} We refer to the systematic uncertainty for
$\sum_{\substack {i,j}}\epsilon_{ij}Br_{i}Br_{j}Br(D^{*-}\to\pi^-\bar{D}^0)$
and $(\sum_{\substack {i,j}}\epsilon^a_{ij}Br_{i}Br_{j}Br(D^{*0}\to\pi^0D^0)+
\sum_{\substack {i,j}}\epsilon^b_{ij}Br_{i}Br_{j}Br(D^{*-}\to\pi^0D^-))$
as the efficiency related systematic uncertainty for $\pi^+D^0\bar{D}^0$-tagged
and $\pi^+D^-D^0$-tagged processes, respectively.
The efficiency related systematic uncertainty includes the uncertainties from MC statistics, PID, tracking,
$\pi^0$ and $K^0_S$ reconstruction, kinematic fit, cross feed and branching fractions of $D$ and $D^{*}$ decay.
The uncertainty due to finite MC statistics is taken as the uncertainty of the signal efficiency.
A systematic uncertainty of 1\% is assigned to each track for the difference between data and simulation in tracking or PID~\cite{PRL112-022001}.
For $\pi^0$ reconstruction, the corresponding uncertainty is 3\% per $\pi^0$~\cite{PRD81-052005}.
For $K^0_S$ reconstruction, the corresponding uncertainty is 4\% per $K^0_S$~\cite{PRD87-052005}.
The uncertainty due to the kinematic fit is estimated by applying the track-parameter corrections to the track helix parameters
and the corresponding covariance matrix for all charged tracks to obtain improved agreement between data and MC
simulation~\cite{PRD87-012002}.
The difference between the obtained efficiencies with and without this correction is taken as the systematic uncertainty for the kinematic fit.
The cross feed among different decay modes is estimated using the signal MC simulation
as detailed in Tables~\ref{tab:sys_eff_D0D0}--\ref{tab:sys_eff_D0Dm2} of Appendix~\ref{sec:appB}. 
The systematic uncertainties for the branching fractions of $D$ and $D^{*}$ decay are estimated by PDG~\cite{pdg}.
A summary of the systematic uncertainties for signal efficiency is listed in
Tables~\ref{tab:sys_eff_D0D0}--\ref{tab:sys_eff_D0Dm2} of Appendix~\ref{sec:appB}.
The total efficiency related systematic uncertainties are combined
by considering the correlation of uncertainties between each decay channels.

\begin{table*}[hbtp]
\begin{center}   
\caption{Summary of systematic uncertainties on the pole mass and pole width of the $Z_c(3885)^-$,
and the product of Born cross section times the $(D\bar{D}^*)^-$ branching
fraction of $Z_c(3885)^-$ ($\sigma\times Br$).
The items noted with * are common uncertainties, and other items are independent uncertainties. 
}
\ \\
\begin{tabular}{lcccccccc}
\hline
\hline
\multirow{3}{*}{Source}
& \multirow{2}{*}{$\Delta M_{\rm{pole}}$}
& \multirow{2}{*}{$\Delta\Gamma_{\rm{pole}}$}
& \multicolumn{4}{c}{$\frac{\Delta(\sigma\times Br)}{\sigma\times Br}$ (\%)}
\\
\cline{4-7}
&
& 
& \multicolumn{2}{c}{$\pi^+D^0\bar{D}^0$-tagged process}
& \multicolumn{2}{c}{$\pi^+D^-D^0$-tagged process}
\\
\cline{2-7}
& (MeV/$c^2$)
& (MeV)
& 4.23~GeV
& 4.26~GeV
& 4.23~GeV
& 4.26~GeV
\\
\hline
Beam Energy                    &  1.0  &  1.6  &  3.3 & 3.0  &  4.9 & 3.4  \\
Mass calibration               &  1.0  &       &             &             \\
$\mathcal{L}(1+\delta^{r})(1+\delta^{v})$*     &       &       &  4.7 & 4.7  &  4.7 & 4.7  \\   
Signal shape                   &  0.1  &  0.1  &  0.1 & 0.1  &  0.1 & 0.1  \\
$Z_c(4020)^-$ Signal           &  0.4  &  1.0  &  2.9 & 2.0  &  2.8 & 3.9  \\
Background shape               &  0.4  &  0.1  &  2.0 & 0.5  &  2.9 & 0.9  \\
Fit bias                       &  0.2  &  0.1  &  0.5 & 0.3  &  0.1 & 0.8  \\
Signal region of DT            &  0.2  &  0.7  &  4.2 & 1.4  &  0.8 & 1.4  \\
Efficiency related             &       &       &  8.3 & 8.3  &  7.9 & 7.9  \\
\hline                                                                     
Total                          &  1.6  &  2.1  &  11.5 & 10.3 & 11.2 & 10.7  \\
\hline
\hline
\end{tabular}
\label{table:syserr}
\end{center}
\end{table*} 

%% file: summary.tex
\section{Summary}
\label{sec:summary}
\vspace{-0.4cm}

In summary, based on the data samples of 1092~pb$^{-1}$ taken at $\sqrt{s}$=4.23~GeV
and 826~pb$^{-1}$ taken at $\sqrt{s}$=4.26~GeV,
we perform a study of the process $e^+e^-\to\pi^{-}(D\bar{D}^*)^{+}$
and confirm the existence of the charged charmoniumlike state $Z_c(3885)^{-}$ in the $(D\bar{D}^*)^-$ system.
The angular distribution of the $\pi^{+}-Z_c(3885)^{-}$ system
is consistent with the expectation from a $J^P=1^+$ quantum number assignment.
We perform a simultaneous fit to the $(D\bar{D}^*)^{-}$ mass spectra for
the two isospin processes of $e^+e^-\to\pi^+D^0D^{*-}$
and $e^+e^-\to\pi^+D^-D^{*0}$ using a mass-dependent Breit Wigner function.
The statistical significance of the $Z_c(3885)$ signal is greater than
$10\sigma$. 
The pole mass and pole width of $Z_c(3885)^-$
are determined to be $M_{\rm{pole}}$=(3881.7$\pm$1.6(stat.)$\pm$1.6(syst.))~MeV/$c^2$
and $\Gamma_{\rm{pole}}$=(26.6$\pm$2.0(stat.)$\pm$2.1(syst.))~MeV, respectively.
The products of Born cross section and the $D\bar{D}^*$ branching fraction of $Z_c(3885)^-$
for $e^+e^-\to\pi^+D^0D^{*-}$ and $e^+e^-\to\pi^+D^-D^{*0}$ are combined into a weighted average~\cite{NIMPRSA-346}.
For the data samples at $\sqrt{s}$=4.23~GeV, the result is
$\sigma(e^+e^-\to\pi^{\pm}Z_c(3885)^{\mp})\times Br(Z_c(3885)^{\mp}\to (DD^*)^{\mp})$
= (141.6$\pm$7.9(stat.)$\pm$12.3(syst.))~pb. For the $\sqrt{s}$=4.26~GeV data sample, the result is
$\sigma(e^+e^-\to\pi^{\pm}Z_c(3885)^{\mp})\times Br(Z_c(3885)^{\mp}\to (DD^*)^{\mp})$
= (108.4$\pm$6.9(stat.)$\pm$8.8(syst.))~pb.

The pole mass and pole width of $Z_c(3885)^-$ and
$\sigma(e^+e^-\to\pi^{\pm}Z_c(3885)^{\mp})\times Br(Z_c(3885)^{\mp}\to (DD^*)^{\mp})$
are consistent with but more precise than those of BESIII's previous results~\cite{PRL112-022001},
with significantly improved systematic uncertainties.
The improvement in the results obtained in this analysis is due to the fact that the double $D$ tag technique
and more $D$ tag modes are used
and two isospin processes $e^+e^-\to\pi^{-}(D\bar{D}^*)^{+}$ are fitted simultaneously
with datasets at $\sqrt{s}$ = 4.23 and 4.26~GeV.
This analysis only has $\sim$9\% events in common with the ST analysis~\cite{PRL112-022001},
so the two analyses are almost statistically independent and
can be combined into a weighted average~\cite{comb}.
The combined pole mass and width are $M_{\rm{pole}}=(3882.2\pm1.1
\mathrm{(stat.)} \pm 1.5 \mathrm{(syst.)})\;\mathrm{MeV}/c^2$
and $\Gamma_{\rm{pole}}=(26.5\pm1.7\mathrm{(stat.)}\pm2.1\mathrm{(syst.)})\;\mathrm{MeV}$, respectively.
The combined $\sigma(e^+e^-\to\pi^{\pm}Z_c(3885)^{\mp})\times Br(Z_c(3885)^{\mp}\to (DD^*)^{\mp})$
is $(104.4\pm4.8\mathrm{(stat.)}\pm8.4\mathrm{(syst.)})\;\mathrm{pb}$ at $\sqrt{s}$=4.26~GeV.

%% file: acknowledgement.tex
\begin{acknowledgements}
\label{sec:acknowledgement}
\vspace{-0.4cm}

The BESIII collaboration thanks the staff of BEPCII
and the IHEP computing center for their strong support.
This work is supported in part by National Key Basic Research Program
of China under Contract No.~2015CB856700;
National Natural Science Foundation of China (NSFC) under Contracts
Nos.~10935007, 11075174, 11121092,11125525, 11235011, 11322544, 11335008, 11425524, 11475185;
the Chinese Academy of Sciences (CAS) Large-Scale Scientific Facility Program;
the CAS Center for Excellence in Particle Physics (CCEPP);
the Collaborative Innovation Center for Particles and Interactions (CICPI);
Joint Large-Scale Scientific Facility Funds of the NSFC and CAS under Contracts
Nos.~11179007, U1232201, U1332201;
CAS under Contracts Nos.~KJCX2-YW-N29, KJCX2-YW-N45;
100 Talents Program of CAS;
National 1000 Talents Program of China;
INPAC and Shanghai Key Laboratory for Particle Physics and Cosmology;
German Research Foundation DFG under Contract No.~Collaborative Research Center CRC-1044;
Istituto Nazionale di Fisica Nucleare, Italy;
Ministry of Development of Turkey under Contract No.~DPT2006K-120470;
Russian Foundation for Basic Research under Contract No.~14-07-91152;
The Swedish Resarch Council;
U.S.~Department of Energy under Contracts
Nos.~DE-FG02-04ER41291, DE-FG02-05ER41374, DE-SC0012069, DESC0010118;
U.S.~National Science Foundation;
University of Groningen (RuG) and the Helmholtzzentrum fuer Schwerionenforschung GmbH (GSI), Darmstadt;
WCU Program of National Research Foundation of Korea under Contract No.~R32-2008-000-10155-0.

\end{acknowledgements}

%% file: appA.tex
\begin{appendix}
\section{Signal Efficiency}
\label{sec:appA}
\vspace{-0.4cm}

The signal efficiency for $\pi^+D^0\bar{D}^0$-tagged process at $\sqrt{s}$=4.23~GeV
and $\sqrt{s}$=4.26~GeV are listed Table~\ref{tab:eff_D0D0},
while the signal efficiency for $\pi^+D^-D^0$-tagged process and its isospin channel are listed in Table~\ref{tab:eff_D0Dma} and~\ref{tab:eff_D0Dmb}.

\begin{table*}[!]
\begin{center}
\vspace{-0.4cm}
\caption{Signal efficiency $\epsilon_{ij}$ (\%) for $\pi^+Z_c(3885)^-(Z_c(3885)^-\to D^0D^{*-}),~D^{*-} \to \pi^- \bar{D}^0,~D^0 \to i,~\bar{D}^0 \to j$, where $i$ and $j$ denote the neutral $D$ final states: $K^-\pi^+$, $K^-\pi^+\pi^0$,
$K^-\pi^+\pi^+\pi^-$ and $K^-\pi^+\pi^+\pi^-\pi^0$ (labeled as 0, 1, 2, 3, respectively).
}
\begin{tabular}{ccccccccc}
\hline
\hline
\multirow{2}{*}{\{$i$, $j$\}}
& \multicolumn{2}{c}{0} 
& \multicolumn{2}{c}{1} 
& \multicolumn{2}{c}{2} 
& \multicolumn{2}{c}{3}
\\
\cline{2-9}
& 4.23~GeV
& 4.26~GeV
& 4.23~GeV
& 4.26~GeV
& 4.23~GeV
& 4.26~GeV
& 4.23~GeV
& 4.26~GeV
\\
\hline
0  &  30.23$\pm$0.17 & 30.30$\pm$0.17  & 14.68$\pm$0.12 & 14.76$\pm$0.12  & 17.54$\pm$0.13 & 17.53$\pm$0.13  & 6.50$\pm$0.08 & 6.46$\pm$0.08   \\
1  &  15.23$\pm$0.12 & 15.47$\pm$0.12  &  6.65$\pm$0.08 &  6.52$\pm$0.08  &  7.80$\pm$0.09 &  7.80$\pm$0.09  & 2.45$\pm$0.05 & 2.33$\pm$0.05   \\
2  &  17.42$\pm$0.13 & 17.33$\pm$0.13  &  7.50$\pm$0.09 &  7.45$\pm$0.09  &  8.01$\pm$0.09 &  8.00$\pm$0.09  & 2.30$\pm$0.05 & 2.30$\pm$0.05   \\
3  &   6.64$\pm$0.08 &  6.62$\pm$0.08  &  2.26$\pm$0.05 &  2.29$\pm$0.05  &  2.41$\pm$0.05 &  2.30$\pm$0.05  & 0.35$\pm$0.02 & 0.30$\pm$0.02   \\
\hline
\hline
\end{tabular}
\label{tab:eff_D0D0}
\end{center}
\end{table*}

\begin{table*}[!htbp]
\begin{center}
\caption{Signal efficiencies $\epsilon^{\alpha}_{ij}$ for $\pi^+Z_c(3885)^-(Z_c(3885)^-\to D^-D^{*0}),~D^{*0}\to \pi^0D^0,~D^-\to i,~D^0 \to j$, where $i$ denotes the charged $D$ final states:
$K^+\pi^-\pi^-$, $K^+\pi^-\pi^-\pi^0$, $K^0_S\pi^-$, $K^0_S\pi^-\pi^0$,
$K^0_S\pi^+\pi^-\pi^-$ and $K^+K^-\pi^-$ (labeled as $A,~B,~C,~D,~E$ and $F$, respectively),
and $j$ denotes the neutral $D$ final states: $K^-\pi^+$, $K^-\pi^+\pi^0$,
$K^-\pi^+\pi^+\pi^-$ and $K^-\pi^+\pi^+\pi^-\pi^0$ (labeled as 0, 1, 2, 3, respectively).
}
\begin{tabular}{ccccccccc}
\hline
\hline
\multirow{2}{*}{\{$i$, $j$\}}
& \multicolumn{2}{c}{0} 
& \multicolumn{2}{c}{1} 
& \multicolumn{2}{c}{2} 
& \multicolumn{2}{c}{3}
\\
\cline{2-9}
& 4.23~GeV
& 4.26~GeV
& 4.23~GeV
& 4.26~GeV
& 4.23~GeV
& 4.26~GeV
& 4.23~GeV
& 4.26~GeV
\\
\hline
$A$ & 24.29$\pm$0.16  & 23.96$\pm$0.15  & 11.49$\pm$0.11  & 11.63$\pm$0.11  & 13.61$\pm$0.12  & 13.57$\pm$0.12  & 4.76$\pm$0.07  & 4.58$\pm$0.07   \\
$B$ & 10.78$\pm$0.10  & 10.72$\pm$0.10  &  4.44$\pm$0.07  &  4.44$\pm$0.07  &  4.92$\pm$0.07  &  4.89$\pm$0.07  & 1.21$\pm$0.03  & 1.14$\pm$0.03   \\
$C$ & 24.66$\pm$0.16  & 25.11$\pm$0.16  & 12.02$\pm$0.11  & 12.05$\pm$0.11  & 14.22$\pm$0.12  & 14.27$\pm$0.12  & 5.09$\pm$0.07  & 4.89$\pm$0.07   \\
$D$ & 11.56$\pm$0.11  & 11.55$\pm$0.11  &  4.85$\pm$0.07  &  4.87$\pm$0.07  &  5.79$\pm$0.08  &  5.62$\pm$0.07  & 1.61$\pm$0.04  & 1.53$\pm$0.04   \\
$E$ & 14.56$\pm$0.12  & 14.75$\pm$0.12  &  6.23$\pm$0.08  &  6.31$\pm$0.08  &  6.31$\pm$0.08  &  6.24$\pm$0.08  & 1.70$\pm$0.04  & 1.59$\pm$0.04   \\
$F$ & 19.29$\pm$0.14  & 19.13$\pm$0.14  &  9.05$\pm$0.10  &  9.11$\pm$0.10  & 10.67$\pm$0.10  & 10.64$\pm$0.10  & 3.51$\pm$0.06  & 3.38$\pm$0.06   \\
\hline
\hline
\end{tabular}
\label{tab:eff_D0Dma}
\end{center}
\end{table*}

\begin{table*}[!htbp]
\begin{center}
\caption{Signal efficiencies $\epsilon^{\beta}_{ij}$ for $\pi^+Z_c(3885)^-(Z_c(3885)^-\to D^0D^{*-}),~D^{*-}\to \pi^0D^-,~D^-\to i,~D^0 \to j$,
where $i$ and $j$ are described in the caption of Table~\ref{tab:eff_D0Dma}.
}
\begin{tabular}{ccccccccc}
\hline
\hline
\multirow{2}{*}{\{$i$, $j$\}}
& \multicolumn{2}{c}{0} 
& \multicolumn{2}{c}{1} 
& \multicolumn{2}{c}{2} 
& \multicolumn{2}{c}{3}
\\
\cline{2-9}
& 4.23~GeV
& 4.26~GeV
& 4.23~GeV
& 4.26~GeV
& 4.23~GeV
& 4.26~GeV
& 4.23~GeV
& 4.26~GeV
\\
\hline
$A$ & 23.57$\pm$0.15  & 23.65$\pm$0.15   & 11.32$\pm$0.11   &  11.42$\pm$0.11  & 13.22$\pm$0.11   & 13.09$\pm$0.11   & 4.75$\pm$0.07   & 4.68$\pm$0.07   \\
$B$ & 10.83$\pm$0.10  & 10.49$\pm$0.10   &  4.34$\pm$0.07   &   4.34$\pm$0.07  &  4.86$\pm$0.07   &  4.76$\pm$0.07   & 1.17$\pm$0.03   & 1.16$\pm$0.03   \\
$C$ & 24.51$\pm$0.16  & 24.37$\pm$0.16   & 11.94$\pm$0.11   &  11.91$\pm$0.11  & 13.98$\pm$0.12   & 13.87$\pm$0.12   & 4.96$\pm$0.07   & 4.93$\pm$0.07   \\
$D$ & 11.34$\pm$0.11  & 11.30$\pm$0.11   &  4.68$\pm$0.07   &   4.83$\pm$0.07  &  5.67$\pm$0.08   &  5.46$\pm$0.07   & 1.58$\pm$0.04   & 1.47$\pm$0.04   \\
$E$ & 14.04$\pm$0.12  & 14.17$\pm$0.12   &  6.19$\pm$0.08   &   6.04$\pm$0.08  &  6.11$\pm$0.08   &  6.08$\pm$0.08   & 1.60$\pm$0.04   & 1.52$\pm$0.04   \\
$F$ & 18.89$\pm$0.14  & 18.79$\pm$0.14   &  9.03$\pm$0.10   &   9.08$\pm$0.10  & 10.42$\pm$0.10   & 10.37$\pm$0.10   & 3.35$\pm$0.06   & 3.44$\pm$0.06   \\
\hline
\hline
\end{tabular}
\label{tab:eff_D0Dmb}
\end{center}
\end{table*}

%% file: appB.tex
\section{The Efficiency Related Systematic Uncertainty}
\label{sec:appB}
\vspace{-0.4cm}

The systematic uncertainties for signal efficiency are listed in
Table~\ref{tab:sys_eff_D0D0}--\ref{tab:sys_eff_D0Dm2}.

\begin{table*}[hbtp]
\begin{center}   
\caption{The systematic uncertainties for signal efficiency (\%) for
$\pi^+Z_c(3885)^-(Z_c(3885)^-\to D^0D^{*-}),~D^{*-} \to \pi^- \bar{D}^0,~D^0 \to i,~\bar{D}^0 \to j$,
where $i$ and $j$ are described in the caption of Table~\ref{tab:eff_D0D0}.
}
\ \\
\begin{tabular}{lccccccccccc}
\hline
\hline
\multirow{2}{*}{\{$i$, $j$\}}
& \multirow{2}{*}{PID}
& \multirow{2}{*}{Tracking}
& \multirow{2}{*}{$\pi^0$}
& \multicolumn{2}{c}{Kinematic fit}
& \multicolumn{2}{c}{MC statistics}
& \multicolumn{2}{c}{Cross feed}
& \multicolumn{2}{c}{Total}
\\
\cline{5-12}
&
&
&
& 4.23~GeV
& 4.26~GeV
& 4.23~GeV
& 4.26~GeV
& 4.23~GeV
& 4.26~GeV
& 4.23~GeV
& 4.26~GeV
\\
\hline                                                                                             
\{0, 0\}  &  4   &   5      &   0     &   0.6 & 0.5   &  0.6  & 0.6   &    0.2 & 0.2  &      6.5 & 6.5      \\
\{0, 1\}  &  4   &   5      &   3     &   0.6 & 0.3   &  0.8  & 0.8   &    0.1 & 0.1  &      7.1 & 7.1     \\
\{0, 2\}  &  6   &   7      &   0     &   0.7 & 1.2   &  0.8  & 0.8   &    0.1 & 0.3  &      9.3 & 9.3     \\
\{0, 3\}  &  6   &   7      &   3     &   1.2 & 0.9   &  1.2  & 1.2   &    0.2 & 0.0  &      9.8 & 9.8     \\
\{1, 0\}  &  4   &   5      &   3     &   0.5 & 0.6   &  0.8  & 0.8   &    0.1 & 0.2  &      7.1 & 7.1     \\
\{1, 1\}  &  4   &   5      &   6     &   0.7 & 0.5   &  1.2  & 1.2   &    0.1 & 0.0  &      8.9 & 8.9      \\
\{1, 2\}  &  6   &   7      &   3     &   0.9 & 0.4   &  1.2  & 1.2   &    0.2 & 0.1  &      9.8 & 9.8     \\
\{1, 3\}  &  6   &   7      &   6     &   0.8 & 0.6   &  2.1  & 2.1   &    0.1 & 0.0  &     11.2 & 11.2     \\
\{2, 0\}  &  6   &   7      &   0     &   0.7 & 0.8   &  0.8  & 0.8   &    0.2 & 0.1  &      9.3 & 9.3     \\
\{2, 1\}  &  6   &   7      &   3     &   0.6 & 0.5   &  1.1  & 1.1   &    0.1 & 0.1  &      9.8 & 9.8      \\
\{2, 2\}  &  8   &   9      &   0     &   1.3 & 1.1   &  1.1  & 1.1   &    0.0 & 0.0  &     12.2 & 12.1     \\
\{2, 3\}  &  8   &   9      &   3     &   0.5 & 1.1   &  2.0  & 2.1   &    2.0 & 2.9  &     12.7 & 13.0     \\
\{3, 0\}  &  6   &   7      &   3     &   0.8 & 0.6   &  1.2  & 1.2   &    0.1 & 0.3  &      9.8 & 9.8     \\
\{3, 1\}  &  6   &   7      &   6     &   0.6 & 0.9   &  2.0  & 2.1   &    0.0 & 0.1  &     11.2 & 11.2     \\
\{3, 2\}  &  8   &   9      &   3     &   1.0 & 1.6   &  2.1  & 2.1   &    2.4 & 2.5  &     12.8 & 12.9     \\
\{3, 3\}  &  8   &   9      &   6     &   0.9 & 1.0   &  5.4  & 5.8   &    0.0 & 0.0  &     14.5 & 14.7     \\
\hline                                                    
\hline
\end{tabular}
\label{tab:sys_eff_D0D0}
\end{center}
\end{table*} 

\begin{table*}[hbtp]
\begin{center}   
\caption{The systematic uncertainties for signal efficiency (\%) for
$\pi^+Z_c(3885)^-(Z_c(3885)^-\to D^-D^{*0}),~D^{*0} \to \pi^0 D^0,~D^- \to i,~D^0 \to j$,
where $i$ and $j$ are described in the caption of Table~\ref{tab:eff_D0Dma}.
}
\ \\
\begin{tabular}{lcccccccccccc}
\hline
\hline
\multirow{2}{*}{\{$i$, $j$\}}
& \multirow{2}{*}{PID}
& \multirow{2}{*}{Tracking}
& \multirow{2}{*}{$\pi^0$}
& \multirow{2}{*}{$K^0_S$}
& \multicolumn{2}{c}{Kinematic fit}
& \multicolumn{2}{c}{MC statistics}
& \multicolumn{2}{c}{Cross feed}
& \multicolumn{2}{c}{Total}
\\
\cline{6-13}
&
&
&
&
& 4.23~GeV
& 4.26~GeV
& 4.23~GeV
& 4.26~GeV
& 4.23~GeV
& 4.26~GeV
& 4.23~GeV
& 4.26~GeV
\\
\hline                                                                                                      
\{$A$, 0\}  &  5   &   6      &    0     &  0     &  0.3 & 0.4       &   0.6 & 0.6      &     0.3 & 0.4     &     7.9 & 7.9     \\
\{$B$, 0\}  &  5   &   6      &    3     &  0     &  0.1 & 0.1       &   1.0 & 1.0      &     0.3 & 0.2     &     8.4 & 8.4     \\
\{$C$, 0\}  &  3   &   4      &    0     &  4     &  0.2 & 0.2       &   0.6 & 0.6      &     0.4 & 0.3     &     6.5 & 6.4     \\
\{$D$, 0\}  &  3   &   4      &    3     &  4     &  0.4 & 0.3       &   0.9 & 0.9      &     0.2 & 0.2     &     7.1 & 7.1     \\
\{$E$, 0\}  &  5   &   6      &    0     &  4     &  0.7 & 0.5       &   0.8 & 0.8      &     0.1 & 0.1     &     8.8 & 8.8     \\
\{$F$, 0\}  &  5   &   6      &    0     &  0     &  0.4 & 0.3       &   0.7 & 0.7      &     0.5 & 0.5     &     7.9 & 7.9     \\
\{$A$, 1\}  &  5   &   6      &    3     &  0     &  0.3 & 0.6       &   0.9 & 0.9      &     0.1 & 0.1     &     8.4 & 8.4     \\
\{$B$, 1\}  &  5   &   6      &    6     &  0     &  0.3 & 0.7       &   1.5 & 1.5      &     0.1 & 0.1     &    10.0 & 10.0   \\
\{$C$, 1\}  &  3   &   4      &    3     &  4     &  0.4 & 0.3       &   0.9 & 0.9      &     0.2 & 0.2     &     7.1 & 7.1     \\
\{$D$, 1\}  &  3   &   4      &    6     &  4     &  0.2 & 0.1       &   1.4 & 1.4      &     0.2 & 0.1     &     8.9 & 8.9     \\
\{$E$, 1\}  &  5   &   6      &    3     &  4     &  0.9 & 0.8       &   1.3 & 1.3      &     0.3 & 0.5     &     9.4 & 9.4     \\
\{$F$, 1\}  &  5   &   6      &    3     &  0     &  0.6 & 0.4       &   1.1 & 1.0      &     0.2 & 0.3     &     8.5 & 8.4     \\
\{$A$, 2\}  &  7   &   8      &    0     &  0     &  0.6 & 1.0       &   0.9 & 0.9      &     0.2 & 0.1     &    10.7 & 10.7   \\
\{$B$, 2\}  &  7   &   8      &    3     &  0     &  0.5 & 0.4       &   1.4 & 1.4      &     0.1 & 0.3     &    11.1 & 11.1   \\
\{$C$, 2\}  &  5   &   6      &    0     &  4     &  0.4 & 0.6       &   0.8 & 0.8      &     0.2 & 0.0     &     8.8 & 8.8     \\
\{$D$, 2\}  &  5   &   6      &    3     &  4     &  0.4 & 0.3       &   1.3 & 1.3      &     0.2 & 0.2     &     9.4 & 9.4     \\
\{$E$, 2\}  &  7   &   8      &    0     &  4     &  1.0 & 1.2       &   1.3 & 1.3      &     0.0 & 0.0     &    11.5 & 11.5   \\
\{$F$, 2\}  &  7   &   8      &    0     &  0     &  1.0 & 0.9       &   1.0 & 1.0      &     0.3 & 0.3     &    10.7 & 10.7   \\
\{$A$, 3\}  &  7   &   8      &    3     &  0     &  0.9 & 1.0       &   1.4 & 1.5      &     1.2 & 2.0     &    11.2 & 11.4   \\
\{$B$, 3\}  &  7   &   8      &    6     &  0     &  0.0 & 0.3       &   2.9 & 3.0      &     0.0 & 0.0     &    12.5 & 12.6   \\
\{$C$, 3\}  &  5   &   6      &    3     &  4     &  1.1 & 0.7       &   1.4 & 1.4      &     0.3 & 1.0     &     9.4 & 9.5     \\
\{$D$, 3\}  &  5   &   6      &    6     &  4     &  0.8 & 1.2       &   2.5 & 2.6      &     0.0 & 0.0     &    10.9 & 11.0   \\
\{$E$, 3\}  &  7   &   8      &    3     &  4     &  1.2 & 1.6       &   2.4 & 2.5      &     0.0 & 0.0     &    12.1 & 12.1   \\
\{$F$, 3\}  &  7   &   8      &    3     &  0     &  0.8 & 0.9       &   1.7 & 1.7      &     0.2 & 0.2     &    11.2 & 11.2   \\
\hline
\hline
\end{tabular}
\label{tab:sys_eff_D0Dm1}
\end{center}
\end{table*} 

\begin{table*}[hbtp]
\begin{center}   
\caption{The systematic uncertainties for signal efficiency (\%) for 
$\pi^+Z_c(3885)^-(Z_c(3885)^-\to D^0D^{*-}),~D^{*-} \to \pi^0D^-,~D^- \to i,~D^0 \to j$,
where $i$ and $j$ are described in the caption of Table~\ref{tab:eff_D0Dma}.
}
\ \\ 
\begin{tabular}{lcccccccccccc}
\hline
\hline
\multirow{2}{*}{\{$i$, $j$\}}
& \multirow{2}{*}{PID}
& \multirow{2}{*}{Tracking}
& \multirow{2}{*}{$\pi^0$}
& \multirow{2}{*}{$K^0_S$}
& \multicolumn{2}{c}{Kinematic fit}
& \multicolumn{2}{c}{MC statistics}
& \multicolumn{2}{c}{Cross feed}
& \multicolumn{2}{c}{Total}
\\
\cline{6-13}
&
&
&
&
& 4.23~GeV
& 4.26~GeV
& 4.23~GeV
& 4.26~GeV
& 4.23~GeV
& 4.26~GeV
& 4.23~GeV
& 4.26~GeV
\\
\hline      
\{$A$, 0\}  &  5   &   6      &    0     &  0     &   0.7 & 0.5       &   0.7 & 0.7       &     0.5 & 0.5     &    7.9  & 7.9     \\
\{$B$, 0\}  &  5   &   6      &    3     &  0     &   0.4 & 0.2       &   1.0 & 1.0       &     0.2 & 0.3     &    8.4  & 8.4     \\
\{$C$, 0\}  &  3   &   4      &    0     &  4     &   0.2 & 0.3       &   0.6 & 0.6       &     0.2 & 0.3     &    6.4  & 6.4     \\
\{$D$, 0\}  &  3   &   4      &    3     &  4     &   0.2 & 0.2       &   0.9 & 0.9       &     0.3 & 0.3     &    7.1  & 7.1     \\
\{$E$, 0\}  &  5   &   6      &    0     &  4     &   1.0 & 0.8       &   0.8 & 0.8       &     0.3 & 0.3     &    8.9  & 8.9     \\
\{$F$, 0\}  &  5   &   6      &    0     &  0     &   0.4 & 0.5       &   0.7 & 0.7       &     0.4 & 0.5     &    7.9  & 7.9     \\
\{$A$, 1\}  &  5   &   6      &    3     &  0     &   0.8 & 0.6       &   0.9 & 0.9       &     0.2 & 0.2     &    8.5  & 8.4     \\
\{$B$, 1\}  &  5   &   6      &    6     &  0     &   0.5 & 0.3       &   1.5 & 1.5       &     0.1 & 0.0     &   10.0  & 10.0     \\
\{$C$, 1\}  &  3   &   4      &    3     &  4     &   0.4 & 0.5       &   0.9 & 0.9       &     0.1 & 0.2     &    7.1  & 7.2     \\
\{$D$, 1\}  &  3   &   4      &    6     &  4     &   0.4 & 0.2       &   1.5 & 1.4       &     0.1 & 0.1     &    8.9  & 8.9     \\
\{$E$, 1\}  &  5   &   6      &    3     &  4     &   0.8 & 0.9       &   1.3 & 1.3       &     0.6 & 0.4     &    9.4  & 9.4     \\
\{$F$, 1\}  &  5   &   6      &    3     &  0     &   0.6 & 0.7       &   1.1 & 1.0       &     0.3 & 0.3     &    8.5  & 8.5     \\
\{$A$, 2\}  &  7   &   8      &    0     &  0     &   0.8 & 0.9       &   0.9 & 0.9       &     0.2 & 0.2     &   10.7  & 10.7     \\
\{$B$, 2\}  &  7   &   8      &    3     &  0     &   1.1 & 0.5       &   1.4 & 1.5       &     0.2 & 0.2     &   11.2  & 11.2     \\
\{$C$, 2\}  &  5   &   6      &    0     &  4     &   0.8 & 0.8       &   0.8 & 0.8       &     0.2 & 0.1     &    8.9  & 8.9     \\
\{$D$, 2\}  &  5   &   6      &    3     &  4     &   0.6 & 0.4       &   1.3 & 1.4       &     0.3 & 0.3     &    9.4  & 9.4     \\
\{$E$, 2\}  &  7   &   8      &    0     &  4     &   1.4 & 1.2       &   1.3 & 1.3       &     0.0 & 0.0     &   11.5  & 11.5     \\
\{$F$, 2\}  &  7   &   8      &    0     &  0     &   1.1 & 1.1       &   1.0 & 1.0       &     0.3 & 0.3     &   10.7  & 10.7     \\
\{$A$, 3\}  &  7   &   8      &    3     &  0     &   1.1 & 1.1       &   1.5 & 1.5       &     1.4 & 1.8     &   11.3  & 11.3     \\
\{$B$, 3\}  &  7   &   8      &    6     &  0     &   1.3 & 0.1       &   2.9 & 2.9       &     0.0 & 0.0     &   12.6  & 12.6     \\
\{$C$, 3\}  &  5   &   6      &    3     &  4     &   0.8 & 0.8       &   1.4 & 1.4       &     0.2 & 0.9     &    9.4  & 9.5     \\
\{$D$, 3\}  &  5   &   6      &    6     &  4     &   0.1 & 0.4       &   2.5 & 2.6       &     0.0 & 0.0     &   10.9  & 11.0     \\
\{$E$, 3\}  &  7   &   8      &    3     &  4     &   1.6 & 1.2       &   2.5 & 2.6       &     0.0 & 0.2     &   12.1  & 12.1     \\
\{$F$, 3\}  &  7   &   8      &    3     &  0     &   0.6 & 1.0       &   1.7 & 1.7       &     0.2 & 0.2     &   11.2  & 11.2     \\
\hline
\hline
\end{tabular}
\label{tab:sys_eff_D0Dm2}
\end{center}
\end{table*} 

\end{appendix}

%% file: draft_v8.0.bbl
\begin{thebibliography}{99}
\bibitem{comb} 
We calculate the combined mean value and combined uncertainty using the
method given in Ref.~\cite{NIMPRSA-346}.
The pole mass and width of two analyses don't have common systematic uncertainties,
while the Born cross section has the common systematic uncertainties from $\mathcal{L}(1+\delta^{r})(1+\delta^{v})$.
\end{thebibliography}
